\documentclass{aastex631}

\newcommand{\Q}[1]{\textcolor{blue}{\bf #1}}

\usepackage{subfigure}
\usepackage{rotating}
\usepackage{ulem}
\begin{document}

\title{Inactive longitude and superflare in the active single-lined pre-main sequence binary V2279 Cyg}

\author[0009-0001-9789-3858]{Xueying Hu}
\affiliation{School of Physics and Astronomy, Beijing Normal University, Beijing 100875, People's Republic of China}
\affiliation{Institute for Frontiers in Astronomy and Astrophysics, Beijing Normal University, Beijing 102206, People's Republic of China}

\author[0000-0003-3816-7335]{Tianqi Cang}
\affiliation{School of Physics and Astronomy, Beijing Normal University, Beijing 100875, People's Republic of China}

\author{Jian-Ning Fu}
\affiliation{School of Physics and Astronomy, Beijing Normal University, Beijing 100875, People's Republic of China}
\affiliation{Institute for Frontiers in Astronomy and Astrophysics, Beijing Normal University, Beijing 102206, People's Republic of China}
\affiliation{Xinjiang Astronomical Observatory, Chinese Academy of Sciences, Urumqi 830011, Xinjiang, China}

\author[0000-0002-6457-4607]{Xuan Wang}
\affiliation{School of Physics and Astronomy, Beijing Normal University, Beijing 100875, People's Republic of China}
\affiliation{Institute for Frontiers in Astronomy and Astrophysics, Beijing Normal University, Beijing 102206, People's Republic of China}

\author{Keyu Xing}
\affiliation{School of Physics and Astronomy, Beijing Normal University, Beijing 100875, People's Republic of China}

\author[0009-0004-1774-7167]{Haotian Wang}
\affiliation{School of Physics and Astronomy, Beijing Normal University, Beijing 100875, People's Republic of China}

\author[0000-0001-7624-9222]{Pascal Petit}
\affiliation{Institut de Recherche en Astrophysique et Plan\'etologie, Universit\'e de Toulouse, CNRS, CNES, 14 avenue Edouard Belin, Toulouse, 31400, France}

\author{Jiaxin Wang}
\affiliation{School of Science, Chongqing University of Posts and Telecommunications, Chongqing 400065, China}

\author{Yong Yang}
\affiliation{School of Physical Science and Technology, Kunming University, Kunming 650214, People's Republic of China}

\author[0000-0003-2645-6869]{He Zhao}
\affiliation{Purple Mountain Observatory and Key Laboratory of Radio Astronomy, Chinese Academy of Sciences, 10 Yuanhua Road, Nanjing 210033, China}

\correspondingauthor{Tianqi Cang \& Jian-Ning Fu}\email{tianqi\_cang@bnu.edu.cn \& jnfu@bnu.edu.cn}

\begin{abstract}
Young, solar-like stars in the pre-main sequence (PMS) stage exhibit vigorous magnetic activity that significantly influences their circumstellar environments and the processes of planetary formation and evolution. In binary systems, tidal forces and magnetic interactions can further shape the magnetic geometry. We report a longitudinal preference of star spots, chromospheric activities, and flares in the active single-lined spectroscopic PMS binary system V2279 Cyg, based on long-term photometric observations from \textit{Kepler} and \textit{TESS} alongside spectroscopic data from LAMOST. The system is classified as a weak-line T Tauri binary, with component masses estimated at 0.86 $M_\odot$ and 0.27 $M_\odot$. V2279 Cyg's nearly circular orbit is synchronized with its 4.126-day rotational period. Observations reveal large star spot regions clustered near the far-side hemisphere. Spectroscopic data show strong, double-peak H$\alpha$ emission, the strength of which is highly correlated with star spot distribution, indicating the presence of an active longitude on the primary star. We also mapped the prominence structure co-rotating with the primary star, suggesting a dense structure close to the near-side hemisphere.
Furthermore, we identify an inactive longitude of flares during the 4-year \textit{Kepler} observations, where the frequency of flare activity is significantly reduced after the superior conjunction, marking the first such identification in active binary systems. Additionally, a white light superflare, releasing energy of $2.5 \times 10^{37}$ erg, was detected in \textit{TESS} observations. These findings provide valuable insights into the magnetic field geometry and dynamo processes in PMS binaries, underscoring the critical role of tidal interactions in shaping magnetic activities.
\end{abstract}


\section{Introduction} \label{sec:intro}
Young, solar-like stars in the pre-main sequence (PMS) stage offer opportunities to study the early phases of stellar and planetary evolution. During this stage, stars exhibit strong magnetic activity driven by dynamo processes in their convective envelopes \citep{2014IAUS..302...25H}. This activity manifests in phenomena such as star spots, flares, prominences, and stellar winds, which can significantly influence their circumstellar environments and, by extension, the formation and evolution of planetary systems \citep{2017A&A...597A..14G}. Among PMS stars, T Tauri stars (TTSs) are particularly useful for investigating magnetic activity and its impact on stellar evolution. TTSs, typically a few million years old and with masses ranging from 0.2 to 2 $\mathrm{M_\odot}$, are categorized into two subclasses: classical T Tauri stars (cTTSs), which exhibit significant accretion from circumstellar disks, and weak-line T Tauri stars (wTTSs), which show minimal to no accretion activity and have largely dissipated their surrounding circumstellar material.

The activity level of a star is related to its rotation. In general, the more rapidly the star rotates, the higher the magnetic activity can be observed. However, when the rotation is rapid enough, the activity level reaches the upper limit and does not increase with the more rapid rotation. This so-called saturated regime has been highlighted for PMS stars with X-ray \citep{2016A&A...589A.113A}, and other active stars (e.g., \cite{2014MNRAS.441.2361V}). Theoretical attempts and simulations on this phenomenon remain flawed \citep{2019ApJ...880....6G, 2022ApJ...926...40W}.

In wTTSs, magnetic activity is primarily influenced by rapid stellar rotation and dynamo processes. These stars often exhibit strong magnetic fields, with strengths reaching kilo-Gauss levels in systems like V410 Tau \citep{2010MNRAS.403..159S,2012A&A...548A..95C}, V830 Tau \citep{2015MNRAS.453.3706D}, LkCa 4 \citep{2023MNRAS.520.3049F}, and TWA 6 \& 8A \citep{2019MNRAS.484.5810H}. This magnetic activity gives rise to rotationally modulated light curves due to star spots and frequent flaring events (e.g., \citealt{2004A&A...427..263F}). Components in close binaries also perform rapid rotation, leading to significant stellar activities \citep{1998MNRAS.296..150G}. Observations of magnetic activity in close binaries suggest that spots can sometimes cluster at specific stellar longitudes or the so-called active longitudes, leading to a preference for certain regions of enhanced activity \citep{1998A&A...334..863B}. For a tidally locked system, this longitude preference is bound to the rotational phase \citep{2006Ap&SS.304..145O,2007A&A...461.1057T}, and tends to appear at sub- and anti-stellar points \citep{2024MNRAS.529.4442S}. The dynamo processes responsible for generating these stable, non-axisymmetric fields are influenced by rapid rotation, which is sustained through tidal interactions in binary systems \citep{1997A&A...321..151M, 2005LRSP....2....8B}.

Active longitudes are highly related to the region where flares occur on the Sun \citep{2016ApJ...818..127G}. However, the distribution of flare events in close binary systems is random \citep{2020ApJ...892...58H}. The understanding of energy transfer between magnetic field structure and flares, especially considering the magnetic field and tidal interactions for close binaries, is still rough. Moreover, the enhanced stellar activity in close binaries \citep{2019A&A...626A..22J,2021A&A...647A..62O} may influence the formation and habitability of planets, and contribute to the magnetic evolution of close-in giant planets.

\object[V2279 Cyg]{V2279 Cyg} ($\alpha_{2000} = 19:18:54.46,~\delta_{2000} = +43:49:25.88$) is an intriguing binary system for investigating longitude preference in a young, active environment. Initially classified as a classical Cepheid variable \citep{2000AJ....119.1901A} and later as an RS CVn-type binary \citep{2000IBVS.4898....1D}. V2279 Cyg has been extensively monitored by space-based photometric missions, including \textit{Kepler} mission \citep{2010ApJ...713L..79K} and \textit{Transiting Exoplanet Survey Satellite} (\textit{TESS}, \citealt{2015JATIS...1a4003R}). 
According to multi-color photometric and high-resolution spectroscopic analyses by \citet{2011MNRAS.413.2709S}, V2279 Cyg has different performance from a Cepheid, especially in its strong magnetic activity signatures and amplitude ratios in different bands.
These observations reveal strong rotational modulation consistent with star spots and frequent flaring events, providing an opportunity to study the distribution and evolution of magnetic activity in the system. 

In this paper, we analyze photometric and spectroscopic data from V2279 Cyg to investigate the longitude preference of star spots and flares. We identified regions of enhanced magnetic activity and assessed their persistence over time. In Section \ref{sec:obs}, we describe the observations and data reduction. Section \ref{sec:para} focuses on constraining the system's stellar and orbital parameters. In Section \ref{sec:Longit}, we model the spotted region, reconstruct prominence maps, and analyze the longitude preference of star spots and flares. Finally, we discuss their implications in Section \ref{sec:Dis} and summarize our results in Section \ref{sec:Con} .

\section{Observations} \label{sec:obs}
Photometric monitoring data for V2279 Cyg retrieved from the Mikulski Archive for Space Telescopes (MAST\footnote{\url{https://mast.stsci.edu}}), was collected by \textit{Kepler} (KIC 8022670, from 2009 - 2013, \dataset[doi:10.17909/t9488n]{\doi{10.17909/t9488n}}) and \textit{TESS} (TIC 159106204, 2019 - now, \dataset[doi: 10.17909/t9-nmc8-f686]{\doi{10.17909/t9-nmc8-f686}}) mission. We employed \textit{Kepler} long-cadence light curves (LCs; 1800s exposure time) from Quarters 0 to 17, and \textit{TESS} light curves processed with the official Quick-Look Pipeline (QLP; \citealt{2020RNAAS...4..204H,2020RNAAS...4..206H,2021RNAAS...5..234K,2022RNAAS...6..236K}). This \textit{TESS} observations span multiple sectors, but generally belong to 3 epochs: Sector 14 (Jul 18 to Aug 14 2019), Sectors 40 ? 41 and 54 ? 55 (Jun 25 to Aug 20 2021 \& Jul 9 to Sep 1 2022), and Sector 74 - 75 \& 80 (Jan 3 to Feb 12 2024 \& Jun 18 to Jul 14 2024). Time-resolved spectroscopic observations were carried out by the Large Sky Area Multi-Object Fiber Spectroscopic Telescope (LAMOST), as part of the LAMOST-\textit{Kepler/K2} project \citep{2015ApJS..220...19D,2018ApJS..238...30Z,2020ApJS..251...15Z,2020ApJS..251...27W,2020RAA....20..167F}. V2279 Cyg was observed extensively between 2018 and 2020, resulting in a total of 71 medium-resolution ($R \sim 7,500$) spectra from LAMOST Data Releases 6, 7, and 8, with signal-to-noise ratios (S/Ns) greater than 7. The wavelength coverage is divided into the red arm ($\lambda \in [630, 680] ~ \mathrm{nm}$) and the blue arm ($\lambda \in [495, 535] ~ \mathrm{nm}$). 

\begin{figure*}[ht!]
\centering
\begin{subfigure}
\centering
\includegraphics[width=0.9\textwidth]{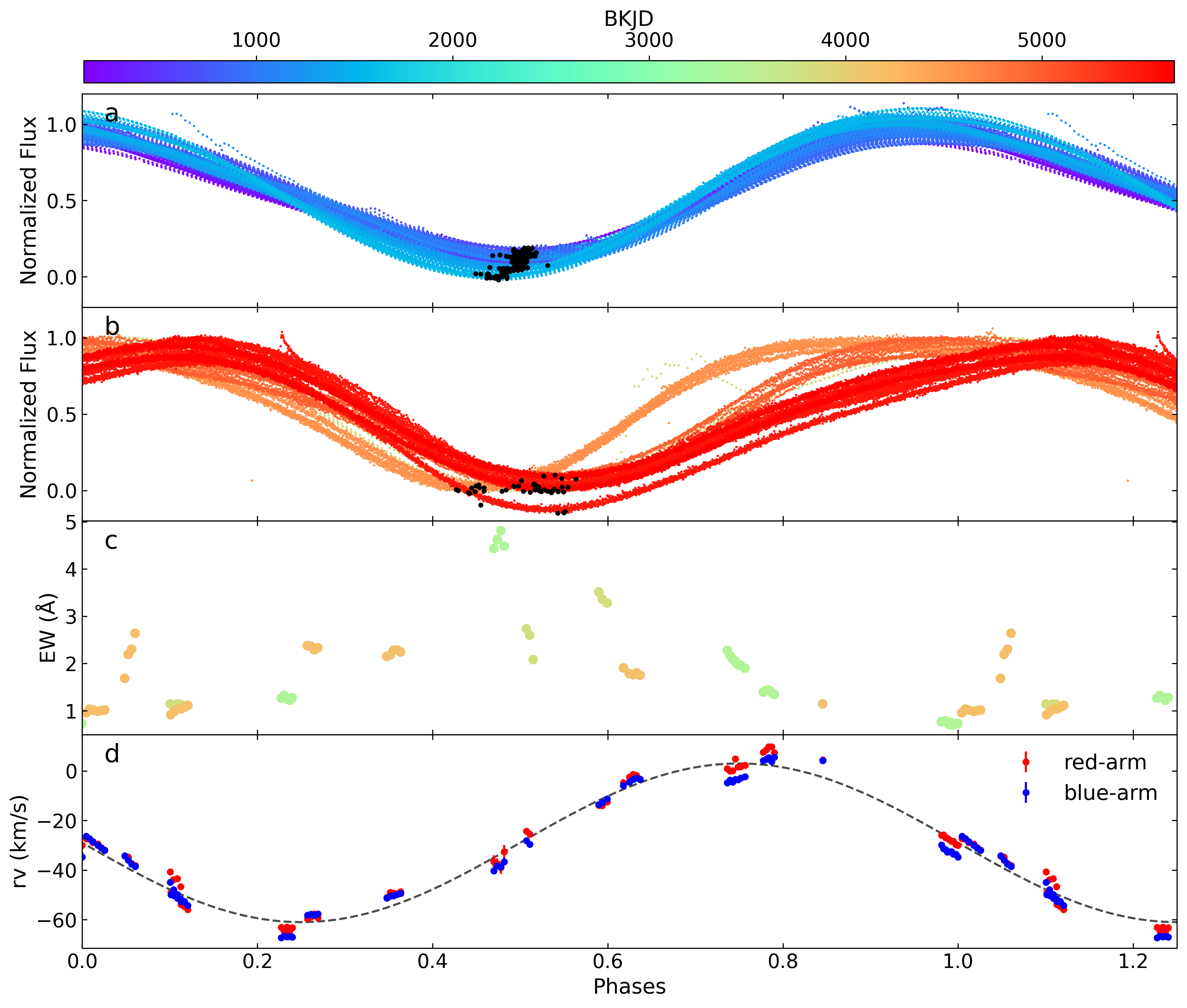}
\end{subfigure}
\caption{The phase-folded curves for V2279 Cyg. (a) \textit{Kepler} light curve. The red dots mark the minimum value in each cycle. (b) \textit{TESS} light curve. The red dots mark the minimum value in each cycle. (c) The equivalent widths (EWs) of H$\alpha$ line for V2279 Cyg from LAMOST MRS. The colorbar for (a-c) represents the Barycentric Kepler Julian Date (BKJD). (d) RV curves of V2279 Cyg, where the red dots represent the RVs measured from the $\mathrm{H\alpha}$ lines of the red-arm spectra ($\mathrm{RV_{r}}$), and the blue dots from the metal lines of the blue-arm spectra ($\mathrm{RV_{b}}$). Table \ref{tab:Rvs} in Appendix \ref{sec:RV} lists the $\mathrm{RV_{r}}$ and $\mathrm{RV_{b}}$. The black dashed line is the sine fitting result using the average value of the data from both arms. 
\label{fig:obs}}
\end{figure*}

The brightness variations observed in these light curves are primarily attributed to rotational modulation caused by star spots \citep{2021A&A...647A..62O}. Using the Lomb-Scargle method, with a signal-to-noise ratio (S/N) threshold of 100, we extracted the primary period of approximately 4.13 days, and a much weaker secondary signal of 2.06 days, which is half the primary one. We note that the rotational modulation pattern in both \textit{Kepler} and \textit{TESS} light curves appears unstable, likely due to the evolution of star spots. This leads to variations in the modulation pattern shape across cycles, as shown in the phase-folded light curves (Fig. \ref{fig:obs}a \& b). 
The variation of pattern can further affect the determination of period since extra frequencies are included to fit the change of light curve shape. In this case, determination and any long-term analysis of the period should consider the variation of the light curve.

Since the light curve shape of Kepler data is stable enough, in order to reduce the effect of light curve shape variation, we conducted a time-frequency analysis using the method described in \citet{2018ApJ...853...98Z}, extracting the frequency around the main period by applying a sliding Lomb?Scargle periodogram (sLSP) analysis with a window width of 200 days and a time step of 5 days. The primary period is relatively stable in the 4-year observation with $P_\mathrm{rot}=4.1264 \pm 0.0001$ days (Appendix. \ref{sec:TF}). Using this period, we derived the following ephemeris:
\begin{equation}
\label{eq:ephm}
\mathrm{BJD} = \mathrm{BJD_0} + 4.1264(1) E,
\end{equation}
where $\mathrm{BJD_0} = 2458268.3132\pm0.0056$ was selected corresponding to the inferior conjunction between two LAMOST observing nights, corrected for light travel time to the solar system's barycenter \citep{2019PASP..131b4506L}, and $E$ the cycle number. 
As shown in Fig. \ref{fig:obs}b, this ephemeris works for \textit{TESS} data, and the patterns in different years are concentrated.

To extract radial velocities (RVs) from the LAMOST spectra, we used the {\tt laspec} package \citep{2021ApJS..256...14Z}. RVs were measured separately for the red- and blue-arm spectra. The folded RV curves, based on the ephemeris derived in Eq.\ref{eq:ephm}, are shown in the bottom panel of Fig. \ref{fig:obs}. The spectroscopic analysis classifies the system as a single-lined spectroscopic binary (SB1). We fitted the RV curve using a single sinusoid function with an RV amplitude of approximately 32 $\mathrm{km~s^{-1}}$, suggesting that the orbital period of the system is nearly identical to the rotational period of the primary star. Additionally, the system's orbital eccentricity is inferred to be close to zero, indicating a nearly circular orbit. 

\begin{table}[]
    \centering
    \caption{Basic stellar parameters of V2279 Cyg from different sources of photometric and spectroscopic observations in optical and infrared. The abbreviation and corresponding references are listed in the references. \label{tab:basic_para}}
    \begin{tabular}{c|ccccccc}
    \hline
         & B20 & F16 & F22 & Z20 & T19 & L19 & S11  \\
        Data source & Board bands & LAMOST &  LAMOST & LAMOST & APOGEE & APOGEE & TREM  \\
    Spec domain & Optical & Optical & Optical & Optical & Infrared & Infrared & Optical  \\
        \hline
        $T_{\mathrm{eff}}$ (K) & 4772 & 4651 & 4985 & 4557 & 4984& 4651 &  4900 \\
        $\mathrm{log}~g$ (dex) &3.256 & 2.96 & 2.64 & 3.07 & 4.388 & 3.10 &3.7 \\
        $[{\rm Fe/H}]$ &-0.043 &-0.08  &-0.06 & -0.262 & -0.391 & -0.5 & -1.2\\
        \hline
    \end{tabular}
\tablerefs{
B20: \cite{2020AJ....159..280B},
F16: \cite{2016A&A...594A..39F},
F22: \cite{2022A&A...664A..78F},
Z20: \cite{2020ApJS..251...15Z},
T19: \cite{2019ApJ...879...69T},
L19: \cite{2019MNRAS.483.3255L},
S11: \cite{2011MNRAS.413.2709S}}
\end{table}

\begin{table}[ht!]

\centering
\caption{System parameters of V2279 Cyg.\label{tab:paras}}
\begin{tabular}{cc|c}
\hline
Parameters & Values & Ref. \\
\hline
Period  (d) & $4.1264 \pm 0.0001$& This work \\
$T_{\mathrm{eff}}$ (K)  & $4651 \pm 115$ & 1\\
$\mathrm{log}~g$ (dex) & $3.10 \pm 0.35$ & 1\\
$[{\rm Fe/H}]$ & $-0.5 \pm 0.1$ & 1\\
$v{\rm sin}~i$ ($\mathrm{km/s^{-1}}$)& $43.2 \pm 3.2$ & 2\\
M1  ($M_{\odot}$) & $0.86 \pm 0.12$ & This work\\
M2  ($M_{\odot}$)& $0.27 \pm 0.03$ & This work\\
Radius (Primary) ($R_{\odot}$)& $3.64 \pm 0.36$ & This work\\
$e$ & $^a~0$ & This work\\
INCL ($^\circ$) & $^a~75$ & This work\\
$R_\mathrm{orb}$ ($R_{\odot}$) & $^a~11.2$ & This work\\
$L$ ($L_{\odot}$) & $5.59 \pm 0.08$ & This work \\
$V$ magnitude ($^\mathrm{mag}$) & 12.7 & 3\\ 
$G$ magnitude ($^\mathrm{mag}$)& 12.6 & 4\\
$A_G$ ($^\mathrm{mag}$) & $^a~0.37$ &This work \\
Distance (pc)& $696 \pm 5$ &4\\
\hline
\end{tabular}
\tablecomments{$^a$ Fixed values}
\tablerefs{
1.\citet{2019MNRAS.483.3255L},
2.\cite{2022A&A...664A..78F},
3.\cite{2009AcA....59...33P},
4.\cite{2020yCat.1350....0G}.}
\end{table}

\section{Stellar parameters and evolutionary stage}\label{sec:para}
We collected measurements of the basic stellar parameters ($T_\mathrm{eff}$, $\mathrm{log}~g$ and [Fe/H]) for V2279 Cyg from seven sources, as summarized in Table \ref{tab:basic_para}. Among these, \citet{2020AJ....159..280B} used broadband photometry, luminosity, and spectroscopic metallicities from Gaia and \textit{Kepler} data, thus providing a reference for the determination of $T_\mathrm{eff}$ and $\mathrm{log}~g$.
All of the measurements indicate that V2279 Cyg is a cool star with $T_\mathrm{eff} < 5000$ K. 

In such temperature range, infrared spectra are less influenced by spotted regions and offer a larger number of lines available for measuring chemical abundances. V2279 Cyg was observed by the APOGEE survey (\citealt{2017AJ....154...94M}, ID: 2M19185446+4349258), which offers high-resolution data ($R \sim 22{,}500$) in infrared band. From APOGEE observations, two independent [Fe/H] measurements point to a low metallicity for V2279 Cyg (-0.391 from \citealt{2019ApJ...879...69T} and -0.5 from \citealt{2019MNRAS.483.3255L}, respectively). However, the $\mathrm{log}~g$ from \citet{2019ApJ...879...69T} is too large to match the luminosity. Meanwhile, ROTFIT pipeline \citep{2016A&A...594A..39F} for LAMOST data tends to overestimate [Fe/H] in metal-poor stars \citep{2022A&A...664A..78F}, which suggests that there could be a systematic bias in the derived $\log g$ for V2279 Cyg.

Additionally, \citet{2011MNRAS.413.2709S} derives an even lower metallicity $[\mathrm{Fe/H}] = -1.2$ from a high-resolution ($R = 48{,}000$, 390?740 nm) TRES spectrum. However, this measurement relies on fitting theoretical template spectra over a relatively limited spectral range, and strong chromospheric emission lines may have affected the fitting. To account for spot effects, we used the {\tt LASP} pipeline \citep{2014IAUS..306..340W} to extract stellar parameters from each LAMOST spectrum acquired at different epochs, instead of the average value from \citet{2020ApJS..251...15Z}. We adopted $T_\mathrm{eff} = 4596 \pm 44$ K, log$ g = 3.149 \pm 0.045$ and [Fe/H]$=-0.217\pm0.029$ from the phase $\phi = 0$, corresponding to the star?s brightest phase (i.e., minimal spot coverage). These values are generally consistent with those reported by \citet{2019MNRAS.483.3255L}, who found $T_\mathrm{eff} = 4651 \pm 115$ K, $\log g = 3.10 \pm 0.35$ and [Fe/H]$=-0.5\pm0.1$. Therefore, we adopt these parameters from APOGEE throughout our subsequent analysis.

Based on the stellar parameters derived above (Table \ref{tab:basic_para}), we estimated the mass of the primary component of V2279 Cyg using the luminosity derived from Gaia astrometric observations \citep{2020AJ....160..252V} and stellar evolutionary models. The absolute magnitude $M_G = 3.02$ in the Gaia G-band was determined using the Gaia parallax $\pi$ and the corresponding apparent G magnitude $m_G$. To account for extinction, we used the monochromatic extinction at 541.4 nm ($A_0$) derived from the recently released 3D dust map by \citet{Dharmawardena2024}\footnote{The dust map is accessible online at \url{https://zenodo.org/records/11448780}.}. The extinction in the G-band was calculated as $A_G = 0.37$, considering the non-linear effects of the stellar spectral energy distribution (SED) and the filter transmission \citep[see detailed discussions in][]{Li2023,MaizApellaniz2024}.

\begin{figure}[ht!]
\includegraphics[width=0.45\linewidth]{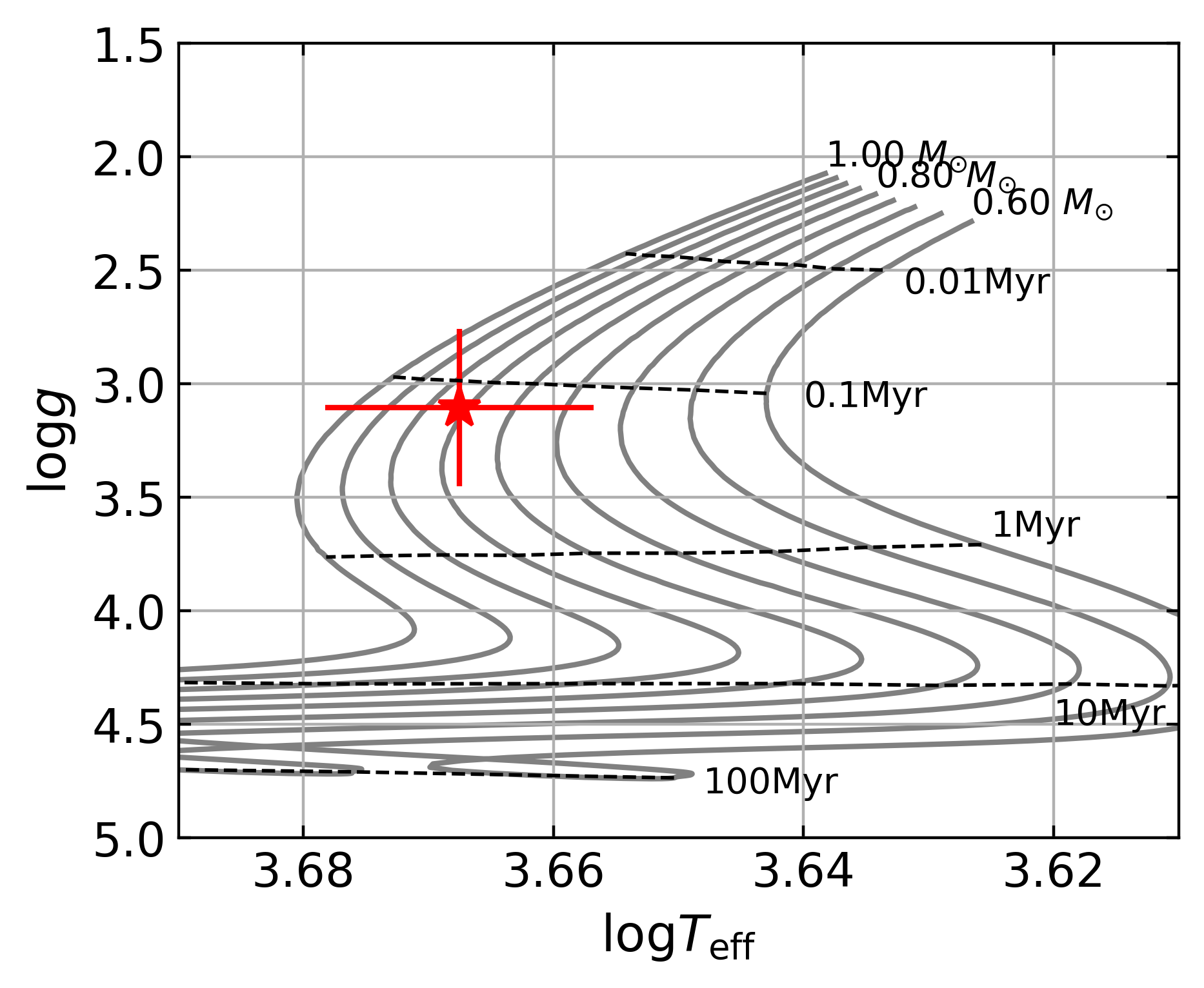}
\includegraphics[width=0.45\linewidth]{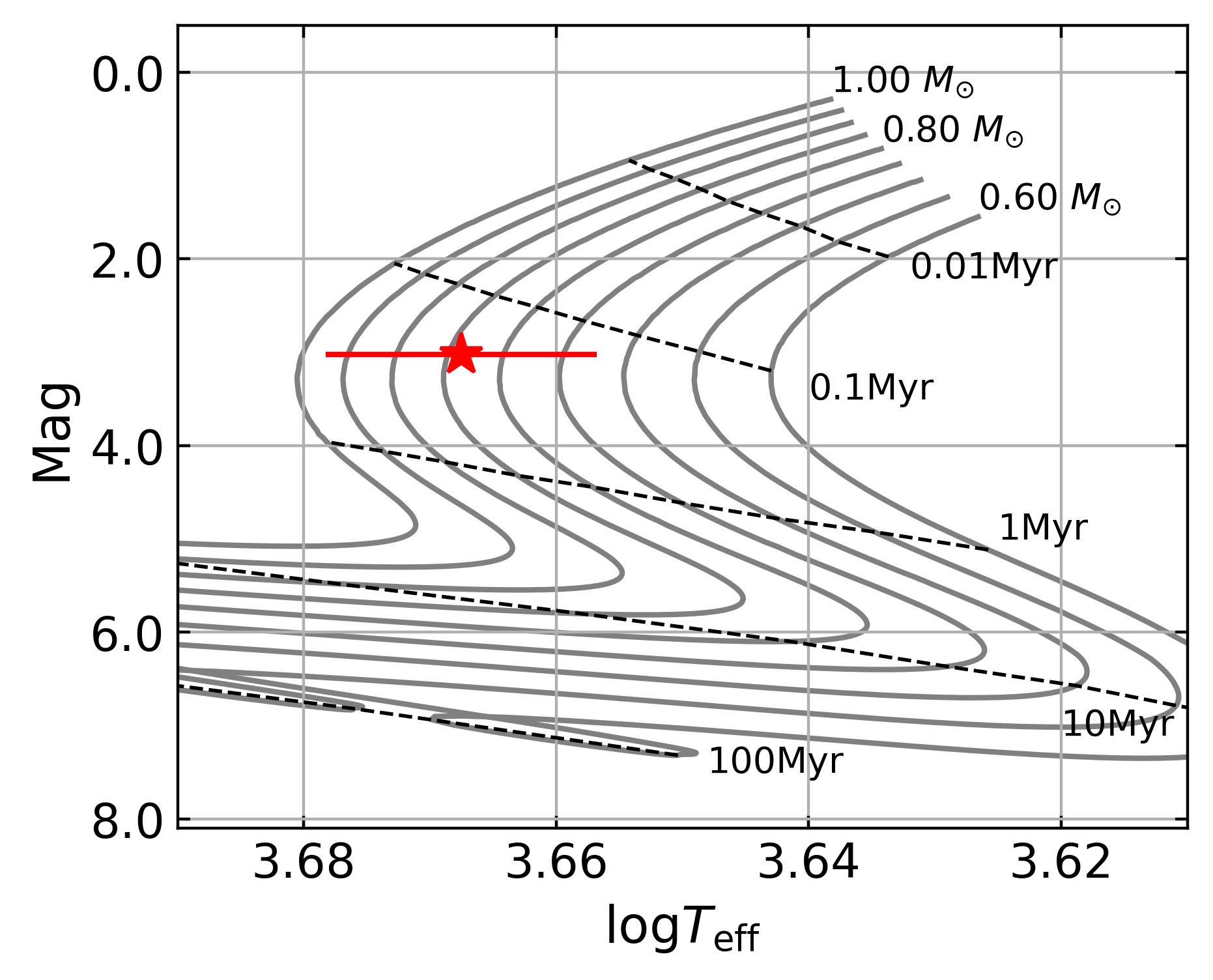}
\caption{Evolutionary tracks from MIST with input masses between 0.6 and 1 $M_{\odot}$ (step: 0.05 $M_{\odot}$) and [Fe/H] of -0.5. The red star marks the position of the primary component with error bars. The black dotted lines are isochrones.}
\label{fig:MIST_track}
\end{figure}

To constrain the mass of the primary star, we employed the MIST\footnote{\url{http://waps.cfa.harvard.edu/MIST/index.html}} stellar evolutionary tracks \citep{2016ApJS..222....8D,2016ApJ...823..102C}, computed with the Modules for Experiments in Stellar Astrophysics (MESA) code \citep{2011ApJS..192....3P,2013ApJS..208....4P,2015ApJS..220...15P}. By comparing $T_\mathrm{eff}$ and $\mathrm{log}~g$ with the evolutionary tracks and using $M_G$ as a constraint, we derived a primary mass of $M_1 = 0.84 \pm 0.12 \mathrm{M_\odot}$ based on $M_G$, and $M_1 = 0.86 \pm 0.12 \mathrm{M_\odot}$ based on $\mathrm{log}~g$ (uncertainty taken from the interpolation of grids). Both results confirm that the primary star is a pre-main sequence (PMS) star, as shown in Fig. \ref{fig:MIST_track}.

As for the secondary component, we adopted the binary mass function, assuming that the eccentricity of the orbit is $e = 0$, we have
\begin{equation}
f = \frac{M_2^3 \sin^3 i}{(M_1 + M_2)^2} = \frac{P_\mathrm K^3}{2 \pi G}.
\end{equation}
A sinusoidal fit to the radial velocity points (Fig. \ref{fig:obs} d) gave the peak radial velocity $K \sim 32 \mathrm{km/s}$. \cite{2022A&A...664A..78F} measured the $v\mathrm{sin}i$ for V2279 Cyg as 43.2 $\mathrm{km/s}$. From $L = 4 \pi R^2 \sigma T^4$, we estimated the radius for the primary star as about $R_1 = 3.64 \pm 0.36 ~\mathrm{R_{\odot}}$  Here, the bolometric luminosity $L_\mathrm{bol} = 5.59 \pm 0.08 ~\mathrm{L_{\odot}}$ is derived by:
\begin{equation}
 -2.5 \mathrm{log} L_\mathrm{bol} = M_{\rm G} + \mathrm{BC_G} - M_\mathrm{bol,\odot},
\end{equation}
where $\mathrm{BC_G}=-0.15$ is estimated by the bolometric correction tool for the Gaia DR3 G-band \citep{2023A&A...674A..26C} with the $T_{\rm eff}$, log $g$, and [Fe/H]. $M_{bol,\odot}=4.74$ is the bolometric luminosity of the Sun \citep{2015arXiv151006262M}. We can then estimate the equatorial velocity $v_{\rm eq} = 44.6$ km/s. Taking the projected equatorial velocity $v{\rm sin}~i = 43.2\pm3.2$ km/s from the LAMOST mid-resolution spectra \citep{2022A&A...664A..78F}, we obtained an estimation of the rotational inclination angle of $i\sim 75^{\circ}$. Thus the mass of the secondary component is $M_2 = 0.27 ~\mathrm{M_{\odot}}$, and the radius of the orbit $R_\mathrm{orb}= 11.3 ~\mathrm{R_{\odot}}$. Since the more massive component does not reach the main sequence, the secondary component should also be a pre-main sequence star. All stellar parameters used in this study are summarized in Table \ref{tab:paras}.

\section{Preference Longitudes of spots and flares} \label{sec:Longit}
\subsection{Spotted region}
Over the \textit{Kepler} observation period (2009 - 2013), the variation in the light curve shape was minor, with the phase of the brightness minimum remaining roughly consistent at around $\phi\sim0.5$ from cycle to cycle as shown in Fig. \ref{fig:obs}. This suggests that the overall brightness distribution on the surface of the primary star was generally stable, i.e., the active longitude was facing the secondary star. For the \textit{TESS} data, the phase-folded light curves exhibit three rotation modulation shapes corresponding to three distinct observation epochs (2019, 2021-22, 2024). Although a bit extended, the brightness minimum is also located between 0.45 and 0.55, keeping the same active longitude as \textit{Kepler} data. A single-peak, continuous modulation pattern appears in all cycles of \textit{Kepler} and \textit{TESS} observations, indicating a large spotted region facing the secondary star that transits the surface with stellar rotating. Further confirmation can be obtained using both the two-spot model and line profile asymmetries. Since the evaluation of the location of the spotted region is sensitive to the rotational phase (or longitude), the preference of the spotted region can be concluded as active longitude, like other binaries.

We examined variations in the spectral line profiles, thanks to the large projected equatorial velocity ($v~\sin i$) of V2279 Cyg, which provides spatial information about the surface brightness distribution \citep{1983PASP...95..565V}. However, due to the limited resolution and S/N of the spectra, direct detection of line profile distortions caused by star spots in individual spectral lines was challenging.

\begin{figure}[ht!]
\centering
\includegraphics[width=0.45\linewidth]{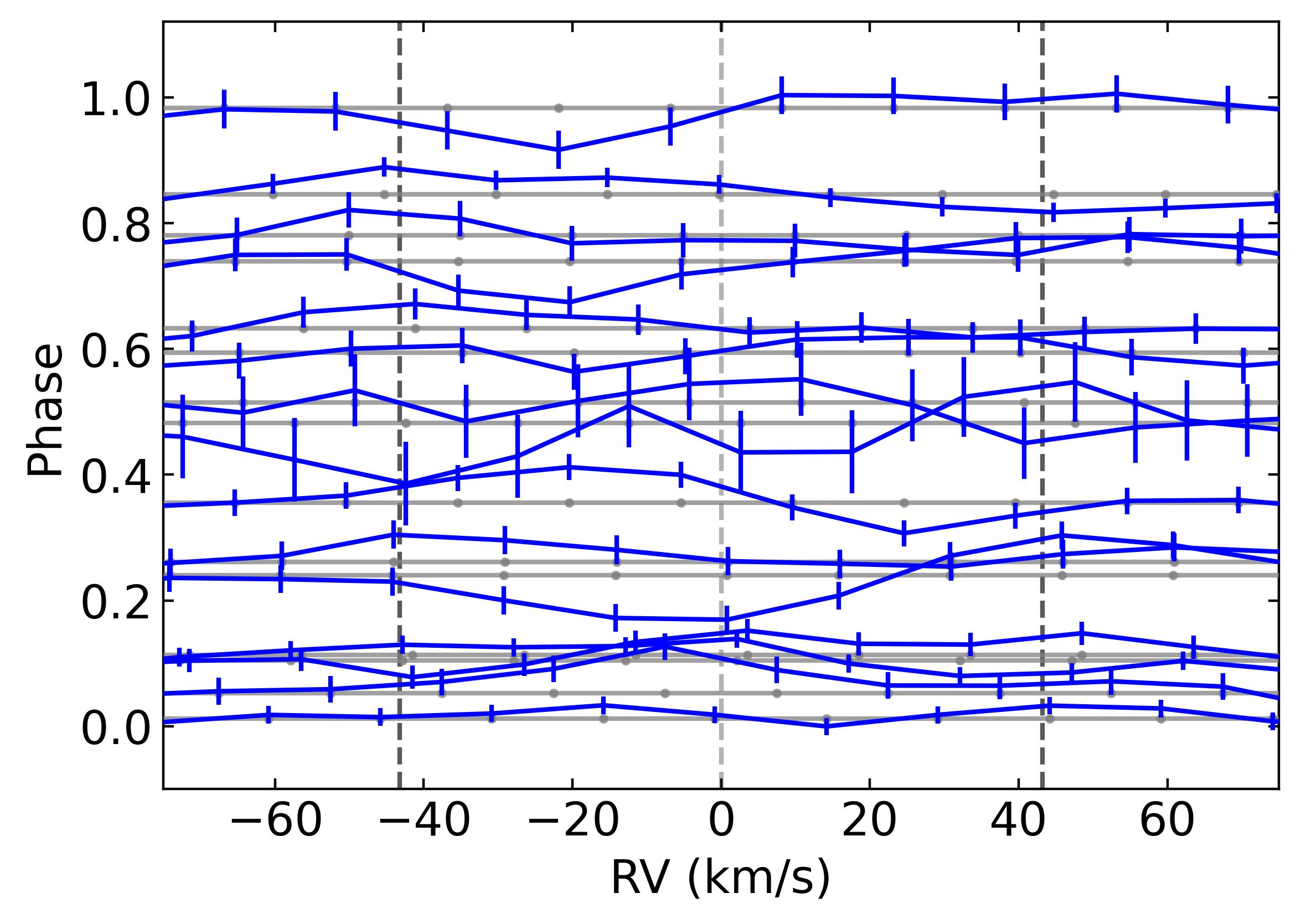}
\caption{Difference between LSD profiles and the Gaussian fitting (gray horizontal lines) of the mean LSD profile. The vertical gray dashed lines mark the $\pm v~\sin i$. Uncertainties (vertical blue lines) are estimated with the LSD process from the observational uncertainty.}
\label{fig:lsd_variation}
\end{figure}

To address this, we employed the Least-Squares Deconvolution (LSD) technique \citep{1997MNRAS.291..658D}, implemented via the Python-based routine LSDpy\footnote{\url{https://github.com/folsomcp/LSDpy}} \citep{2018MNRAS.474.4956F}, to compute average pseudo-line profiles and enhance the S/N. For the LSD analysis, a line mask was constructed using a typical list of photospheric absorption lines (strong chromospheric lines are excluded) for stars with $T_\mathrm{eff} = 4750$ K and $\log g = 3.0$, selected from the Vienna Atomic Line Database (VALD) \citep{1995A&AS..112..525P,1999A&AS..138..119K}. The LSD profiles were computed with a velocity step of 15 $\mathrm{km~s^{-1}}$, slightly smaller than the velocity step of the LAMOST medium-resolution spectra. Figure. \ref{fig:lsd_profile} in Appendix. \ref{sec:ha} shows the LSD profiles for all rotational phases of V2279 Cyg' spectra, corrected for radial velocity. To isolate surface brightness variations, the mean photospheric structure, modeled as a Gaussian profile, was subtracted from all LSD profiles as shown in Fig. \ref{fig:lsd_variation}. A large portion of residual profiles shows an antisymmetric structure with opposite signs on the red and blue sides, indicating the presence of a large spotted region. 

Assuming fixed orbital parameters for the V2279 Cyg binary system as described in Sect. \ref{sec:para}, we modeled the star spots on the surface of the primary component using the PHysics Of Eclipsing BinariEs (PHOEBE) software, version 2.4\footnote{\url{http://phoebe-project.org}} \citep{2005ApJ...628..426P,2016ApJS..227...29P,2018ApJS..237...26H,2020ApJS..247...63J,2020ApJS..250...34C}.

The modeling employed the built-in Markov Chain Monte Carlo (MCMC) method, implemented via the {\tt emcee} package \citep{2013PASP..125..306F}, to fit the light curve and derive a two-spot solution (parameters listed in Table \ref{tab:paras}). For simplicity, we focused on the \textit{Kepler} Quarter 1 (Q1) data, assuming minimal changes in the light curve over time. As shown in Fig. \ref{fig:obs}, the light curve variations in other quarters are minor, and the Q1-derived model is applicable to them as well.

\begin{figure*}
\centering
\includegraphics[width=1\textwidth]{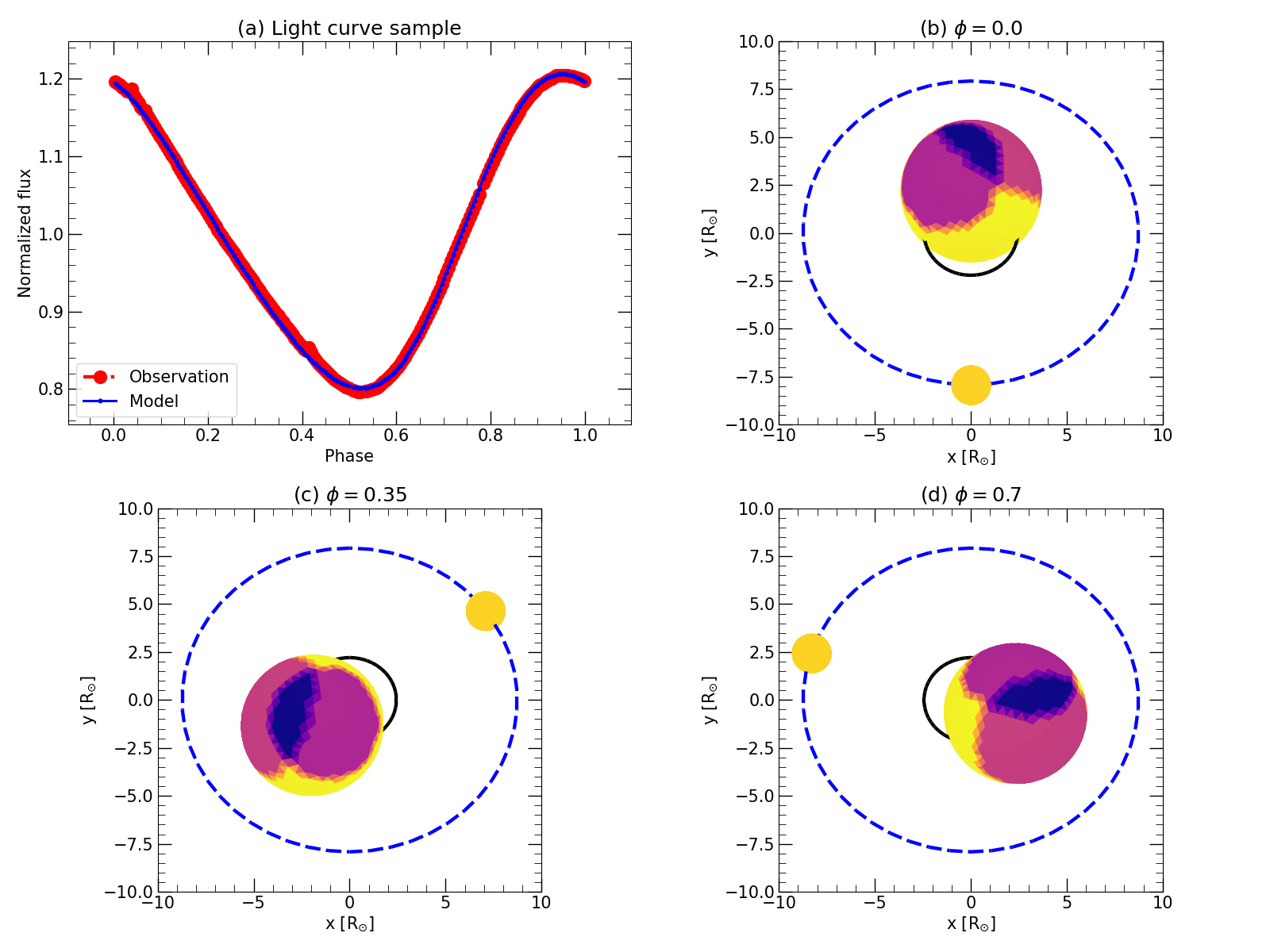}
\caption{The spot model to fit the light curve of \textit{Kepler} Q1 data set. (a) The normalized flux in the phase-folded diagram for the model (blue line) and observations (black dot). (b-d) The configurations of the two-spot model at $\phi$=0, 0.35 and 0.7, respectively. The view illustrates the line-of-sight perspective (considered the inclination) of the two-spot model. Solid lines show the orbit of the primary (black) and secondary (blue) stars.}
\label{fig:bs_model} 
\end{figure*}

The resulting model, shown in Fig. \ref{fig:bs_model}, identifies two large, dark spots with an overlapping region near the pole, creating an even darker area. This configuration reproduces the observed light curve well. The spots remain visible throughout the rotational cycle, consistent with the LSD analysis, which reveals absorption line asymmetries in all the observations.

Although a two-spot model can fit the spot-modulated light curve, which is in agreement with other spot modeling of wTTSs  \citep{2016A&A...592A.140L,2021MNRAS.501.1878X}, the actual brightness distribution can be much more complex, as shown by Doppler imaging (e.g., LkCa 4 by \cite{2023MNRAS.520.3049F} provided an example for fitting photometry and spectroscopic data simultaneously). Furthermore, the two-spot model can be highly impacted by the degeneration of parameters and create spurious parameters \citep{2005MNRAS.359..729J}, which may lead to the difficulty of converging parameters in the MCMC method. There are many local minima in the MCMC sampling in our modeling that walkers converge quickly into one of them (see Fig. \ref{fig:MCMC}). However, this toy model still provides global information that the heavily spotted area covers a large region of the V2279 Cyg surface. The spectral line distortion also indicates that a large spotted region transits the disk, which confirms the conclusion from the light curve.

\subsection{Longitude preference of flares}
\begin{figure}[ht!]
\plotone{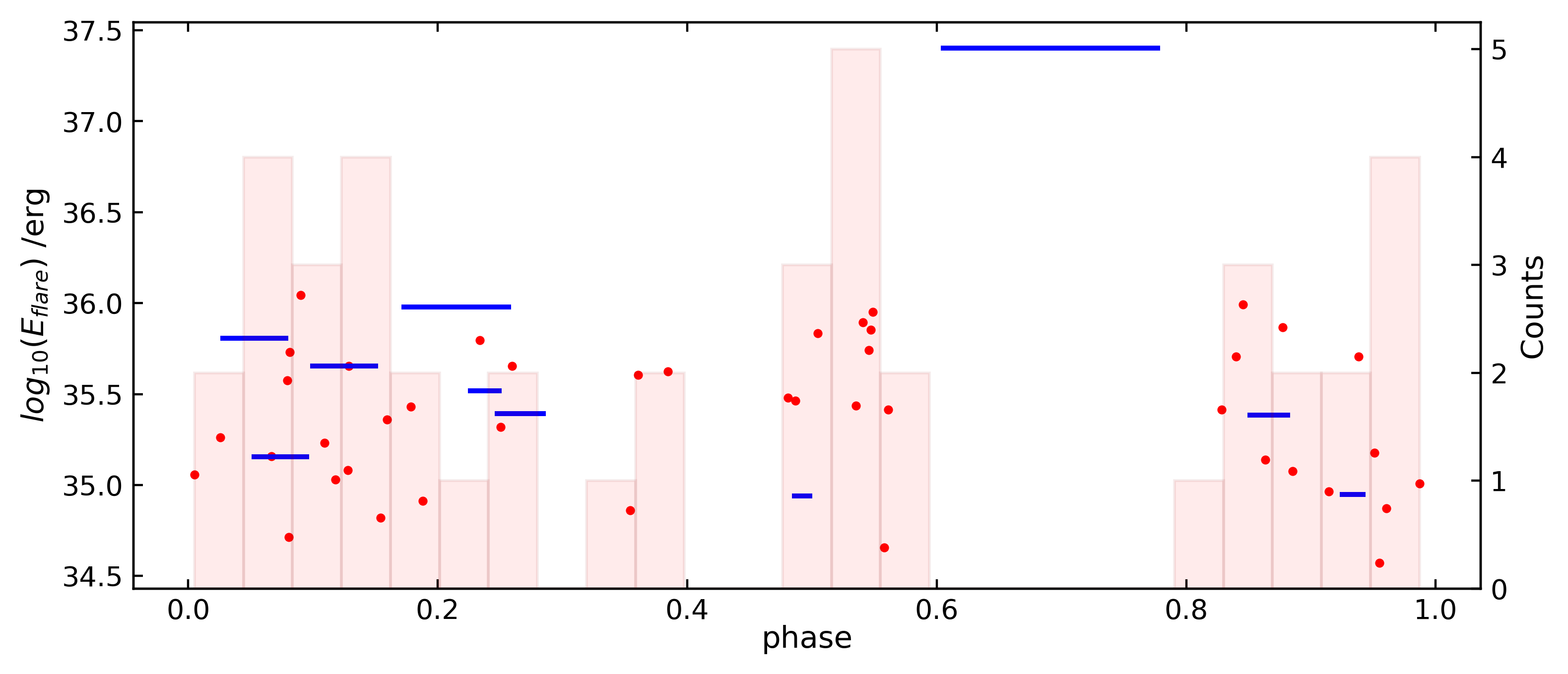}
\caption{The phase-folded flare energies calculated from \textit{Kepler} LC light curves \citep{2022A&A...668A.101O} and \textit{TESS} (our work). The red dots mark the \textit{Kepler} results (at the peak time), and the blue lines represent the \textit{TESS} results (labeling the duration). The light red histogram displays the number of flares at different phases.
\label{fig:flare}}
\end{figure}

The occurrence of flares is also believed to correlate with active regions dominated by star spots for wTTSs (e.g., \citealt{2004A&A...427..263F,2018MNRAS.480.1754N}). A total of 43 flares were identified in the \textit{Kepler} dataset (Quarters 0?17) by \citet{2021A&A...647A..62O,2022A&A...668A.101O}. For the \textit{TESS} data, we employed a flare detection tool designed explicitly for \textit{TESS} light curves, developed by \citet{2024ApJS..271...57X}, and identified 10 additional flares for V2279 Cyg. To estimate the energy of the flares detected in the \textit{TESS} data, we followed the method described in \citet{2018A&A...616A.163V} and used the flare analysis routines provided by \citet{2024ApJS..271...57X}. The method obtained bolometric flare energy by correcting the effect of the \textit{Kepler} and \textit{TESS} passbands, respectively.
The calculated flare energies are listed in Appendix \ref{sec:tessflare}. As shown in the phase-folded flare energy distribution in Fig. \ref{fig:flare}, {\bf NO} flare was detected in the \textit{Kepler} data within the phase range of 0.60 to 0.78, except a superflare observed in \textit{TESS} Sector 14 and covered almost the gap. Assuming that the occurrence of flares is generally random, the 43 \textit{Kepler} flares indicate a likelihood of $<0.03\%$ that no flare would be detected within a 0.18 phase interval during continuous monitoring. In this case, we define this longitude preference of flares as an "inactive longitude". The other \textit{TESS} flares occurred at similar phases to those observed in the \textit{Kepler} light curves, with a likely random distribution. Note that the superflare released an energy of approximately $2.53 \times 10^{37} ~ \mathrm{erg}$, which is at least an order of magnitude larger than the others (details in Appendix \ref{sec:tessflare}).

\subsection{Asymmetric H-alpha emission}
The active longitude is correlated with the chromospheric activity performed by the emission lines in other active binaries (e.g., \citealt{2018MNRAS.480.1754N, 2021MNRAS.504.2461N}). In the red arm of LAMOST medium-resolution spectra, we noticed strong emission as well as profile variations in H$\alpha$ lines. The H$\alpha$ profile was generally the same in one observation run with $1 \sim 9$ spectra in 3 hours from LAMOST, which is much shorter than the 4.13-day orbital period. Therefore, the spectral data from observations in a single night are averaged together to improve the S/N (Fig. \ref{fig:ha_line}). H$\alpha$ lines are purely emitting, sometimes with two peaks slightly outside the $v\mathrm{sin}~i\sim 43$ km/s, e.g., observation on May 30, 2018. In general, the peaks can reach $\pm80$ km/s, and the whole emission structure can extend to about $\pm300$km/s, which is much larger than $v\mathrm{sin}~i$. The H$\alpha$ emission on May 31, 2018, was significantly stronger than that at other times, which can be attributed to a flare. The significant stellar activity events on the primary component of V2279 Cyg suggest a strong magnetic field on the stellar surface that can support prominences like other solar-like stars \citep{2019MNRAS.485.1448V}. Although young binary stellar systems can host circumstellar or circumbinary discs \citep{akeson2019resolved}, we attribute the H$\alpha$ emission of V2279 Cyg to the chromospheric activity due to the ultraviolet excess and no significant infrared excess in SED as shown in Fig. \ref{fig:SED}, where the infrared observations are taken from WISE \citep{2010AJ....140.1868W}.
The enhanced blue peak is also identified in other active wTTSs \citep{2018MNRAS.480.1754N,2021MNRAS.504.2461N}, which can be explained by the mass motion towards the observer. 

The equivalent width (EW) of H$\alpha$ lines shows a weak rotational modulation (Fig. \ref{fig:obs}c). Still, we can recognize a maximum EW at around $\phi=0.5$ and a minimum EW at around $\phi=0$, corresponding to the most spotted region and the brightest region of the star. We calculated the Pearson correlation of -0.78 between the variation of EW and the average phase-folded light curve of \textit{Kepler} data, indicating that a more spotted region has a higher chromospheric activity for V2279 Cyg.

One potential scenario of the asymmetric H$\alpha$ emission in the young active stars without disk can be attributed to the cool, dense gas confined by the strong magnetic field and co-rotating with the stars, or the so-called slingshot prominences \citep{1996IAUS..176..449C, 2021MNRAS.504.1969Z}. This scenario can be applied to V2279 Cyg due to the large inclination $\sim75^\circ$, which results in almost no transit event and absorption features in H$\alpha$ line profiles - the slingshot prominences can always be seen from the line-of-sight. We notice that in the dynamic spectra for 2020 (Fig. \ref{fig:ha_line}), the difference between the line profiles of close phases is minor, with a close position of peaks. For example, the line profile on June 15, 2020, is similar to the one on June 3 after 12 days. We suppose that these repeating profiles indicate a globally stable structure co-rotating with the star in a short-term time duration.

\begin{figure}[ht!]
    \centering
        \includegraphics[width=0.99\linewidth]{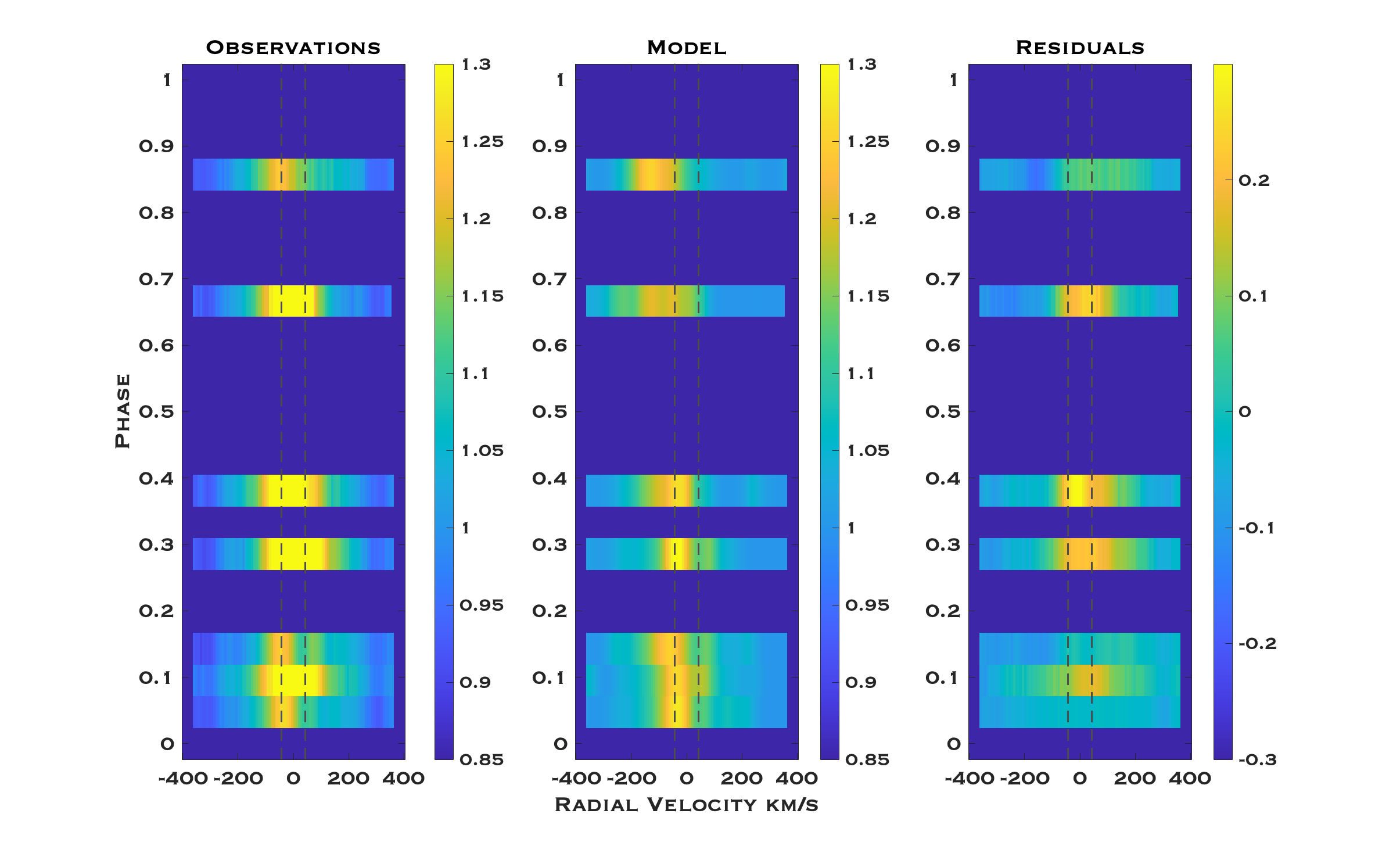} 
    \caption{Dynamic spectra showing the H$\alpha$ line of V2279 Cyg (in 2020). From left to right: observations (after correcting the rv caused by the binary system), model and model residual. The vertical black dashed lines mark the $\pm v \mathrm{sin}i$.}
    \label{fig:DS}
\end{figure}

In order to model the co-rotating structure around the primary component, we employed Doppler tomography \citep{1988MNRAS.235..269M} for spectra in 2020, assuming that the main prominence structure is generally stable in about two weeks and the emitting materials are optically thin. We used the tomography code developed by \citet{2000MNRAS.316..699D}, which is based on the algorithm of \citet{1988MNRAS.235..269M} and was used in other prominence systems \citep{2020A&A...643A..39C, 2005IAUS..226..511P}. The system is assumed to be in a quasi-steady state so that each region of emitting gas maintains a stable velocity in the rotating/orbital frame, and the only change in the observed line profiles (as a function of orbital or rotational phase) is due to the changing line-of-sight. Assuming that the H$\alpha$ emitting materials are optically thin, we used the local H$\alpha$ profile of Gaussian shape with a $\sim 15~\mathrm {km~s^ {-1}}$ Gaussian FWHM, and located them to the 2D velocity map ($V_x$, $V_y$). In the practice of the reconstruction, we neglected the absorption of the star for the detection of hot materials outside the star \citep{2000MNRAS.316..699D}. As shown in the dynamic spectrum in Fig. \ref{fig:DS}, the tomography model (middle panel) roughly fits the pattern of emissions in the observations (left panel), with residuals (right panel) approximately an order of magnitude lower than the observed fluxes. The mismatch mainly happens between $\phi=0.4$ and $\phi=0.7$, where is the most spotted region.

\begin{figure}[ht!]
\centering
        \includegraphics[width=0.6\linewidth]{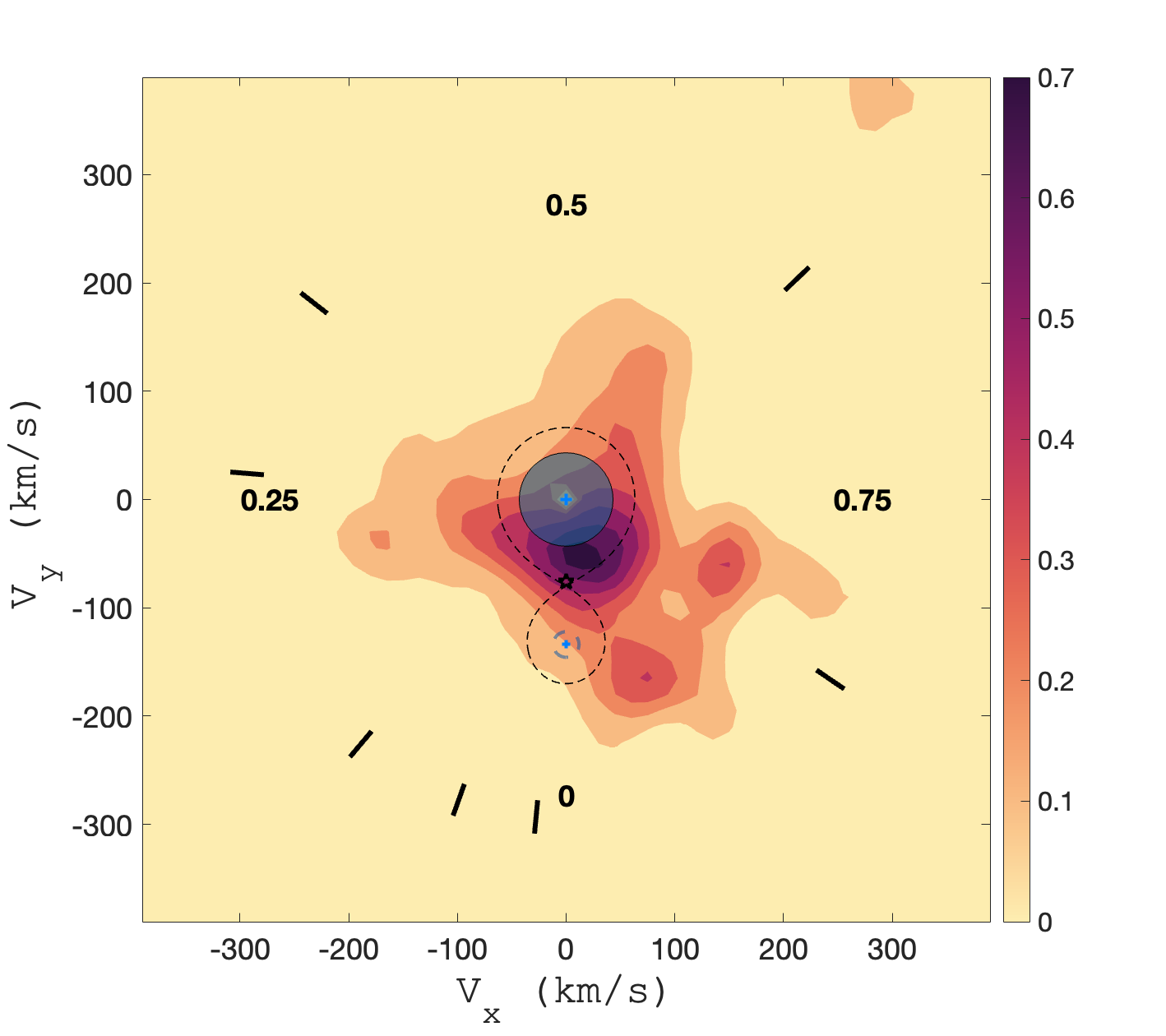} %pp2020_final.png
    \caption{Prominence map (2020). The positions and relative sizes of the primary and secondary stars are marked with filled and dotted blue circles. The dotted black lines indicate the Roche lobe. The short black lines mark the observation phases. The color scale represents the local H$\alpha$ equivalent width, in units of picometers per $\mathrm{15 ~ km ~ s^{-1}}$ square pixel.}
    \label{fig:pro}
\end{figure}

The resulting spatial distribution of prominences is shown in Fig. \ref{fig:pro}, in which the central zero point marks the position of the primary star in the velocity field. In the prominence map, the H$\alpha$ emitting materials concentrate mainly at the phase $\phi \sim 0$, near the surface of the primary star, which is close to the brightness maxima of \textit{Kepler} light curve, near the inferior conjunction. Unfortunately, although 2020 provided the largest number of observations among the three seasons, the data were unevenly distributed across the orbital phases. Furthermore, the limited phase coverage from 2018 and 2019 (only 5 and 3 nights, respectively) made it impossible to reconstruct prominence maps for those years, hindering our ability to discern the long-term structures surrounding the primary star and the binary system.

\section{Discussion} \label{sec:Dis}
\subsection{Evolutionary state of V2279 Cyg}
From the MIST evolutionary tracks, we derived the age of V2279 Cyg as $\sim0.16$ Myr, which is younger than the typical TTSs aged from 0.5?10 Myr, even younger for wTTSs that have dissipated their discs. Since we did not find clues for accretion gas around V2279 Cyg, the abnormally determined age of V2279 Cyg is potentially due to the interaction between the components of the binary system. The mass transition in the earlier stage and the potential accreting materials in the orbit plane may be cleared by the tidal effect between the two components (refer to the review by \citet{1994ARA&A..32..465M}).

The circular orbits and tidal lock found in the V2279 Cyg system widely exist in the other close binary systems with an orbit period of shorter than 7 days \citep{2019ApJ...871..174S}. Although low-mass binaries in the PMS stage are easier to reach synchronization due to the larger size and convective envelope \citep{1989A&A...223..112Z}, V2279 Cyg has a very limited time for angular momentum evolution \citep{2013A&A...556A..36G}, which requires strong tidal torques and magnetic braking to drive the process \citep{2019ApJ...881...88F}. 

\begin{figure}[ht!]
\plotone{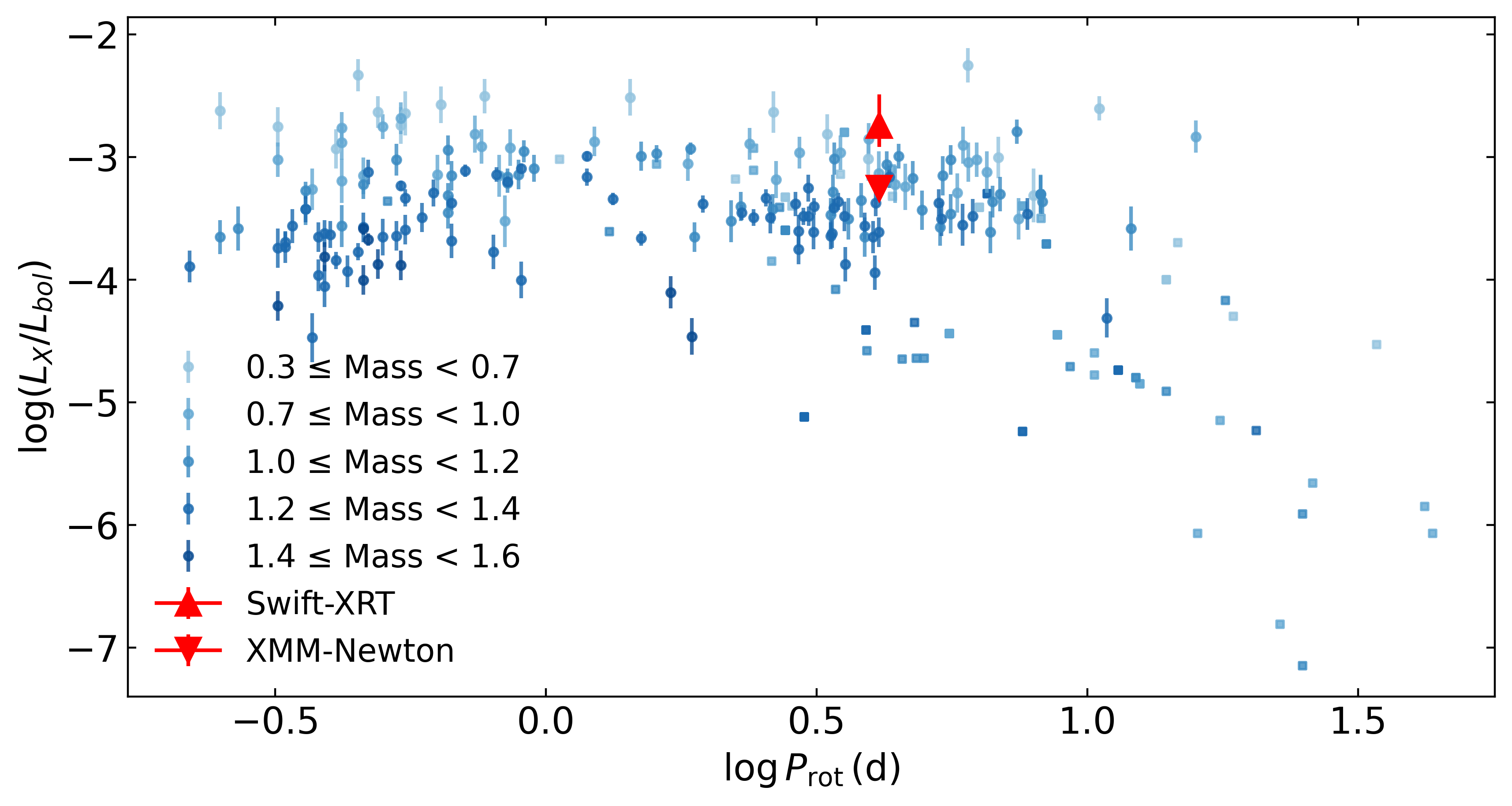}
\caption{The logarithmic fractional X-ray luminosity $\mathrm{log}(L_X/L_{bol})$ vs. rotational period for V2279 Cyg, comparing with h Per members (circles) of different mass \citep{2016A&A...589A.113A}, and MS stars (squares) in the unsaturated and saturated regime from \citet{2014MNRAS.441.2361V}. The flat region on the left with a small period indicates a saturated regime of stellar dynamo. The X-ray emission of V2279 Cyg almost reaches the upper limit of saturated stars.}
\label{fig:Lxp}
\end{figure}

Furthermore, the rapid rotation of V2279 Cyg may lead to a saturated dynamo. As shown in Fig. \ref{fig:Lxp}, we locate V2279 Cyg in the diagram of $L_X/L_\mathrm{bol}$ as a function of the rotation period and compare it with the PMS stars in the young cluster h Persei ($\sim$ 13 Myr old) \citep{2016A&A...589A.113A}. Both measurements from Swift-XRT and XMM-Newton show that V2279 Cyg also lies in the so-called saturation region.

\subsection{Active and inactive longitudes}
The light curve analysis of V2279 Cyg reveals a consistent brightness minimum at $\phi \approx 0.5$, corresponding to a large star spot region around far-side hemisphere that persists across both \textit{Kepler} and \textit{TESS} observations, suggesting a tidally stabilized active longitude, which is also identified in other active binary systems (e.g., \citealt{1996A&A...314..153J,2006Ap&SS.304..145O, 2016A&A...593A.123O}). This phenomenon can be attributed to the constraint of the flux tubes under the influence of tidal effects \citep{2003A&A...405..291H, 2003A&A...405..303H,2005LRSP....2....8B}, with relatively weak differential rotation \citep{1995A&A...294..155M}. The absence of flares around $\phi= 0.60$ to 0.78 during the \textit{Kepler} observations indicates an inactive longitude of flare in this region. Assuming the inclination of $75^\circ$, the active region of the flare region can be roughly constrained in a small circle around latitude $71^\circ$ and $\phi\sim0.2$, with a radius of a few degrees. This area does not show a significant connection to the spotted region, differing from single stars, where flares typically occur at the same longitudes as the most spotted regions (e.g., the Sun \citealt{2000ApJ...540..583S} and wTTSs \citealt{2004A&A...427..263F}). The inactive longitudes of flares in V2279 Cyg are likely due to the magnetic structure being limited by the binary nature, which is similar to the active longitude. On the other hand, the superflare found in the TESS observation suggests a different activity level from the time during \textit{Kepler} observation, leading to a higher occurrence rate of flares in TESS observations as shown in the corresponding cumulative flare frequency distribution (FFD) shown in Fig. \ref{fig:flare_ffd}. Since the number of flares is still small for a solid statistic, more observations would help to investigate the active longitude of V2279 Cyg.

As a PMS, V2279 Cyg is expected to have a deep convective envelope or even be fully convective, similar to M-dwarfs. A recent study on the fully convective M-dwarf close binary CM Dra \citep{2024MNRAS.528..963M} reported no preference for orbital phase. Similarly, well-studied PMS binary system DQ Tau \citep{2018ApJ...862...44K}, with a rotation period of about 16 days, also shows no preferences for flare events. The inactive longitudes appear in V2279 Cyg at $\phi\sim0.7$, yet the star?s large inclination suggests that orbital motion is unlikely to be the main source of any obstruction. Meanwhile, DQ Tau also has a large inclination with a relatively stable light curve shape that can be described by a two-spot model. However, the mass ratio (DQ Tau $q=0.93$) and orbital period could generate different tidal effects, thereby influencing the star?s magnetic structure and flare-driving mechanism. On the other hand, the gap of normal flares shares almost the same phase as the only observed superflare, which released energy comparable to the sum of energy of other flares in one year. This coincidence may be connected to the star's magnetic structure. A detailed topology analysis (e.g., Zeeman Doppler imaging), would help clarify the role of magnetic geometry in modulating flare activity in V2279 Cyg.

\subsection{Slingshot prominences as a source of stellar wind?}
The prominences map is based on a minimum model to trace the emitting clouds co-rotating with the system. Our result showed that the emitting material concentrate opposite to the latitude of the spotted region, between the two components of the binary system, similar to the prominence structure to the active binary V471 Tau \citep{2022MNRAS.513.2893Z}. The failure of the fitting is likely to be caused by the heating of chromospheric activity or rapid evolution of prominences \citep{2021A&A...654A..42C}. At around $\phi\sim0.4$ in 2018 and $\phi\sim0.4 ~\&~ 0.5$ in 2019, as shown in Fig. \ref{fig:ha_line}, the three enhanced emissions also happened around the most spotted region ($\phi={0.5}$), indicating that the heating process likely happened in the same area as spotted region, in agreemen t with the strong correlation between the EW and flux variaion. On the other hand, the red-asymmetric structure of the emission profile suggests that the heated materials move away from the line-of-sight. Regarding that at this time, the spotted region is more or less facing us (Fig. \ref{fig:bs_model}), this scenario can be explained by the return of hot matter to the stellar surface, or corona rain \citep{2023MNRAS.526.1646D}.

The presence of superflares also reflects the confinement of a large amount of material by strong magnetic fields (e.g., \citealt{2024LRSP...21....1K}), which could serve as a potential source of stellar winds \citep{2021LRSP...18....3V}. As the H$\alpha$ lines are fully emitting during our observations, we estimated the mass of emitting materials stored around V2279 Cyg using the method based on \cite{1996MNRAS.281..626S}. We used the median value of the H$\alpha$ EWs from the LAMOST spectra and subtracted the EW measured from the synthetic spectrum as used in Fig. \ref{fig:SED}. We obtained a number density of hydrogen atoms of $\approx 3.9\times10^{15} ~m^{-3}$, and the total mass of prominences is $\approx 6.8\times10^{18}$ kg ($\sim 3.4\times 10^{-12}M_\odot$). 
An enhanced H$\alpha$ emission event was observed on May 31, 2018, during which the EW increased by 3\r{A} in 1 day. With the difference of mass, we can roughly estimate the mass loss as $\Delta \mathrm{M} \approx 3.2 \times 10^{18}$ kg ($\sim 1.6\times 10^{-12}M_\odot$). Considering that there are over 10 flares observed by \textit{Kepler} per year, the mass-loss through prominences eruptions can reach the order of $10^{19}$ kg yr$^{-1}$. 

We also obtained the X-ray flux $F_X$ of $6.7^{+3.9}_{-2.7} \times 10^{-13}$ $\mathrm{erg~s^{-1}~cm^{-2}}$ and $2.05^{+0.15}_{-0.15} \times  10^{-13}$ $\mathrm{erg~s^{-1}~cm^{-2}}$ from Swift-XRT \citep[2SXPS J191854.5+434923]{2020ApJS..247...54E} and XMM-Newton \citep[4XMM J191854.5+434925]{2001A&A...365L...1J}, respectively. Using the average X-ray luminosity of $2.5 \times 10^{31} ~\mathrm{erg~s^{-1}}$, the surface X-ray flux is $F_{X\star}\approx 3\times 10^7~\mathrm{erg~s^{-1}~cm^{-2}}$, corresponding to a coronal temperature of $\approx9$ MK applying the relationship from \citet{2015A&A...578A.129J}. 
Taking the solar mass-loss rate of $\dot{M}_\odot=2\times 10^{-14}M_\odot~\rm{yr}^{-1}$, we estimated a stellar mass-loss rate of $\dot{M}_\star\sim583 \dot{M}_\odot$ according to \citet{2017MNRAS.466.1542S}, which is about $2.3\times 10^{19}~\rm{kg}~\rm{yr}^{-1}$. This value is in agreement with our mass-loss estimation from the prominences, indicating that prominences contribute significantly to the stellar wind of V2279 Cyg, as other active young stars \citep{2021MNRAS.505.5104W}.

\section{Conclusion} \label{sec:Con}
In this study, we have investigated the magnetic activity and longitudinal preferences of the pre-main sequence binary system V2279 Cyg using extensive photometric data from \textit{Kepler} and \textit{TESS}, combined with spectroscopic observations from LAMOST. Our analysis reveals that V2279 Cyg is a synchronized, nearly circular orbit binary with primary and secondary masses of approximately 0.86 $M_\odot$ and 0.27 $M_\odot$, respectively. The consistent clustering of large star spot regions near the far-side hemisphere across multiple observational epochs indicates the presence of a tidally stabilized active longitude on the primary star. Concurrently, we have identified an inactive longitude region where flare activity is notably absent during the \textit{Kepler} observation, marking a novel discovery in the context of PMS binaries. The detection of a rare white light superflare, with an energy release of $2.5 \times 10^{37}$ erg, underscores the system's extreme magnetic activity. Additionally, our estimates of mass loss through slingshot prominences, corroborated by X-ray flux measurements, suggest that such prominences play a significant role in contributing to the stellar wind of V2279 Cyg. These findings provide valuable insights into the magnetic field geometry and dynamo processes in young binary systems, highlighting the crucial influence of tidal interactions in shaping both active and inactive longitudinal regions. Future studies, including detailed magnetic field mapping and continued monitoring of flare activities, will further illuminate the complex magnetic interactions and evolutionary dynamics governing pre-main-sequence binaries like V2279 Cyg.

\begin{acknowledgments}
TC is supported by the LAMOST fellowship as a Youth Researcher, which is supported by the Special Funding for Advanced Users, budgeted and administrated by the Center for Astronomical Mega-Science, Chinese Academy of Sciences (CAMS), and acknowledges funding from the China Postdoctoral Science Foundation (2023M730297). TC acknowledges funding from the support of the National Natural Science Foundation of China (NSFC) through grant 12273002. JNF acknowledges funding from the support of NSFC through grants 12090040/12090042/12427804. This work is supported by the China Manned Space Program with grant No. CMS-CSST-2025-A013, and the Central Guidance for Local Science and Technology Development Fund under No. ZYYD2025QY27. JXW acknowledges the support of NSFC through grants 12203010,  and the Natural Science Foundation of Chongqing (CQCSTC) under grants No. CSTB2023NSCQ-MSX1048. YY acknowledges funding from the support of NSFC through 12403101. This research has made use of the VizieR catalogue access tool, CDS,
Strasbourg, France \citep{10.26093/cds/vizier}. The original description 
of the VizieR service was published in \citet{vizier2000}. Finally, We thank the reviewer for valuable comments and suggestions on this manuscript.  
\end{acknowledgments}

\vspace{5mm}
\facilities{\textit{Kepler}, \textit{TESS}, LAMOST, Swift-XRT, XMM-Newton}

\software{astropy \citep{2013A&A...558A..33A,2018AJ....156..123A,2022ApJ...935..167A}, laspec \citep{2021ApJS..256...14Z}, LSDpy: The GitHub repository for LSDpy is available at folsomcp/LSDpy.
}



\clearpage
\appendix
\section{Time-frequency analysis}
\label{sec:TF}

\begin{figure*}[ht!]
\includegraphics[width=0.95\textwidth]{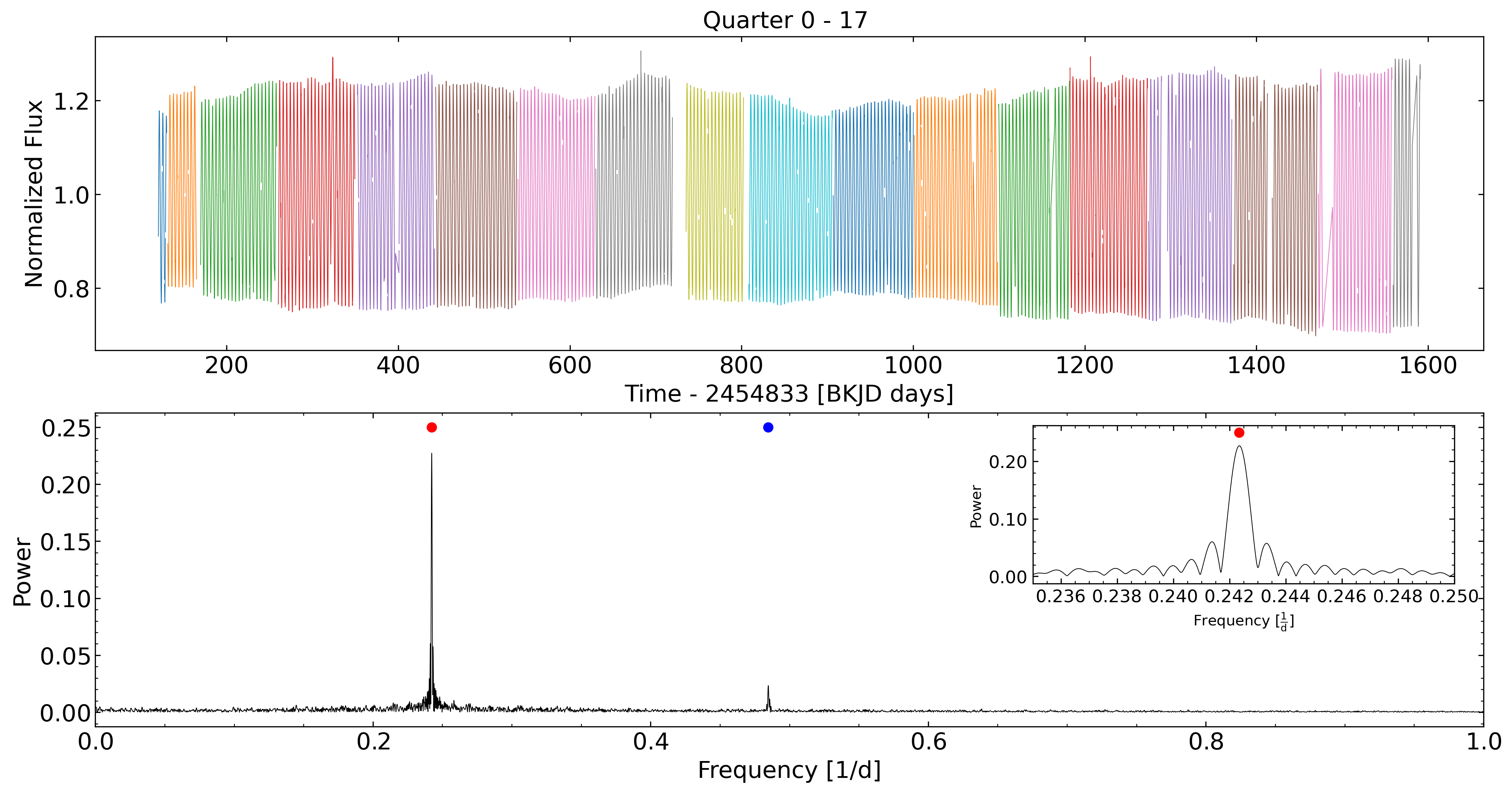}
\caption{\textit{Kepler} LC light curve (upper) and its power spectrum (lower). The red dot in the lower panel marks the main frequency and the blue dot marks its harmonic.
\label{fig:lc_Kepler}}
\end{figure*}

\begin{figure}[ht!]
    \centering
    \includegraphics[width=0.47\textwidth]{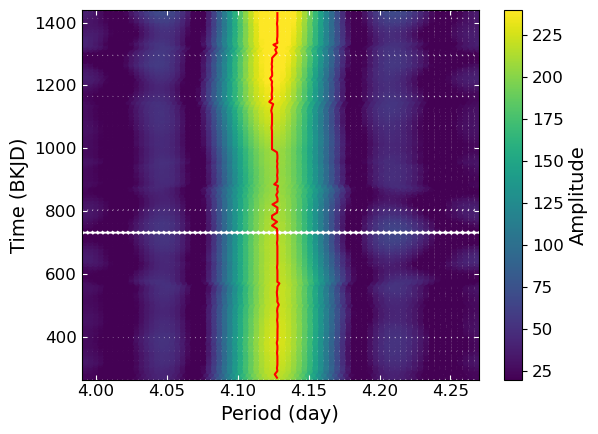}
    \caption{Sliding Lomb?Scargle periodogram (sLSP) for the \textit{Kepler} LC light curve. The red line marks the maximum frequency value.}
    \label{fig:TF}
\end{figure}

We obtained \textit{Kepler} long-cadence light curves Quarter 0-17 and extracted periods of 4.13 and 2.06 days using the Lomb-Scargle method ($S/N > 100$) (Fig. \ref{fig:lc_Kepler}). To analyze the stability of the primary period, we performed a time-frequency analysis on \textit{Kepler} data and found the primary period to be relatively stable in the 4-year observations with $P_\mathrm{rot}=4.1264 \pm 0.0001$ days (Fig. \ref{fig:TF}). The uncertainty is taken from the standard deviation from sLSP.

% -----------------------------
\clearpage
\section{Radial Velocities and EWs\label{sec:RV}}

Table 2 lists the radial velocities of V2279 Cyg measured from both red- and blue-arm spectra of LAMOST MRS, derived with {\tt laspec}, and EWs from H$\alpha$.

\begin{longtable}{rrrrrrr}
\caption{The radial velocities of V2279 Cyg measured from the red- and blue-arm spectra of LAMOST medium-resolution survey. (Column 1 is the BJD time, Column 2 is the corresponding phase, and Columns 3 and 5 are RVs from LAMOST red- and blue-arm, respectively. Columns 4 and 6 are the corresponding errors. Column 7 lists the EWs.)}
\label{tab:Rvs}\\
\hline
\textbf{BJD} & \textbf{Phase} & $\mathbf{RV_r}$ & $\mathbf{err_r}$ & $\mathbf{RV_b}$ & $\mathbf{err_b}$ & \textbf{EW} \\
(day) &  & (km/s) & (km/s) & (km/s) & (km/s) & (A) \\
\hline
\endfirsthead

\hline
\textbf{BJD} & \textbf{Phase} & $\mathbf{RV_r}$ & $\mathbf{err_r}$ & $\mathbf{RV_b}$ & $\mathbf{err_b}$ & \textbf{EW}\\
(day) &  & (km/s) & (km/s) & (km/s) & (km/s) & (A) \\
\hline
\endhead

\hline
\endfoot

\hline
\endlastfoot

2458263.27 & 0.778 & 7.56 & 0.57 & 4.14 & 0.45 & 1.396 \\
2458263.28 & 0.781 & 8.52 & 0.53 & 4.71 & 0.25 & 1.430 \\
2458263.30 & 0.784 & 9.77 & 0.83 & 5.39 & 0.36 & 1.438 \\
2458263.31 & 0.787 & 9.69 & 1.07 & 3.78 & 0.51 & 1.392 \\
2458263.32 & 0.790 & 7.28 & 0.82 & 5.57 & 0.49 & 1.345 \\
2458267.23 & 0.736 & 0.86 & 0.51 & -4.75 & 0.36 & 2.274 \\
2458267.24 & 0.739 & 0.02 & 0.66 & -3.75 & 0.34 & 2.173 \\
2458267.25 & 0.743 & 0.12 & 0.52 & -4.45 & 0.31 & 2.103 \\
2458267.26 & 0.746 & 4.91 & 0.52 & -3.45 & 0.32 & 2.041 \\
2458267.28 & 0.749 & 1.61 & 0.81 & -3.55 & 0.18 & 1.973 \\
2458267.29 & 0.752 & 1.78 & 0.66 & -2.93 & 0.30 & 1.960 \\
2458267.31 & 0.757 & 2.27 & 0.65 & -2.40 & 0.26 & 1.901 \\
2458268.24 & 0.981 & -25.96 & 0.84 & -29.85 & 0.37 & 0.770 \\
2458268.25 & 0.983 & -25.75 & 0.37 & -31.31 & 0.27 & 0.772 \\
2458268.25 & 0.985 & -26.87 & 0.76 & -31.71 & 0.25 & 0.791 \\
2458268.26 & 0.988 & -27.21 & 0.44 & -32.71 & 0.40 & 0.757 \\
2458268.27 & 0.990 & -27.75 & 0.64 & -32.65 & 0.30 & 0.708 \\
2458268.28 & 0.992 & -28.41 & 0.66 & -32.60 & 0.32 & 0.750 \\
2458268.29 & 0.995 & -28.41 & 0.90 & -33.50 & 0.44 & 0.704 \\
2458268.30 & 0.997 & -29.72 & 0.74 & -33.62 & 0.38 & 0.723 \\
2458268.31 & 0.000 & -30.00 & 0.95 & -34.72 & 0.34 & 0.730 \\
2458269.25 & 0.227 & -63.08 & 0.36 & -67.38 & 0.19 & 1.268 \\
2458269.26 & 0.230 & -64.53 & 0.46 & -66.58 & 0.17 & 1.325 \\
2458269.28 & 0.234 & -62.92 & 0.28 & -66.97 & 0.20 & 1.280 \\
2458269.29 & 0.237 & -64.67 & 0.80 & -66.78 & 0.34 & 1.223 \\
2458269.30 & 0.240 & -63.19 & 0.37 & -67.07 & 0.24 & 1.278 \\
2458270.25 & 0.470 & -36.65 & 2.62 & -40.28 & 0.99 & 4.435 \\
2458270.27 & 0.474 & -37.53 & 2.17 & -38.30 & 1.00 & 4.622 \\
2458270.29 & 0.478 & -39.02 & 2.55 & -38.73 & 1.83 & 4.812 \\
2458270.30 & 0.482 & -32.53 & 2.59 & -36.59 & 1.50 & 4.484 \\
2458625.27 & 0.507 & -24.35 & 1.02 & -28.18 & 0.55 & 2.736 \\
2458625.29 & 0.511 & -25.60 & 1.36 & -29.57 & 0.60 & 2.599 \\
2458644.23 & 0.100 & -40.71 & 0.62 & -45.00 & 0.19 & 1.145 \\
2458644.24 & 0.104 & -43.82 & 0.52 & -47.91 & 0.16 & 1.111 \\
2458644.26 & 0.108 & -43.44 & 0.46 & -49.79 & 0.21 & 1.143 \\
2458644.27 & 0.112 & -46.70 & 0.79 & -51.73 & 0.15 & 1.139 \\
2458646.24 & 0.590 & -13.91 & 0.91 & -13.60 & 0.34 & 3.514 \\
2458646.26 & 0.594 & -13.99 & 0.76 & -12.81 & 0.32 & 3.359 \\
2458646.28 & 0.599 & -12.58 & 1.06 & -11.28 & 0.28 & 3.280 \\
2459001.23 & 0.618 & -4.74 & 1.44 & -5.85 & 0.41 & 1.908 \\
2459001.25 & 0.625 & -2.48 & 0.52 & -4.48 & 0.18 & 1.788 \\
2459001.27 & 0.629 & -1.39 & 0.41 & -3.40 & 0.17 & 1.764 \\
2459001.29 & 0.633 & -1.79 & 0.61 & -2.95 & 0.24 & 1.803 \\
2459001.30 & 0.637 & -3.31 & 1.10 & -3.38 & 0.58 & 1.753 \\
2459003.22 & 0.101 & -49.01 & 0.72 & -49.98 & 0.31 & 0.917 \\
2459003.24 & 0.105 & -50.00 & 0.56 & -50.41 & 0.28 & 0.986 \\
2459003.25 & 0.109 & -51.47 & 0.48 & -51.36 & 0.25 & 1.036 \\
2459003.27 & 0.113 & -53.82 & 0.59 & -52.70 & 0.24 & 1.041 \\
2459003.28 & 0.117 & -54.76 & 0.58 & -52.69 & 0.29 & 1.076 \\
2459003.30 & 0.121 & -55.92 & 0.57 & -54.28 & 0.20 & 1.115 \\
2459004.24 & 0.348 & -50.98 & 0.55 & -51.37 & 0.23 & 2.151 \\
2459004.25 & 0.352 & -48.99 & 0.39 & -50.45 & 0.18 & 2.181 \\
2459004.27 & 0.356 & -49.32 & 0.52 & -50.42 & 0.19 & 2.278 \\
2459004.29 & 0.360 & -49.57 & 0.48 & -49.75 & 0.26 & 2.287 \\
2459004.30 & 0.363 & -48.69 & 0.54 & -49.35 & 0.29 & 2.246 \\
2459006.29 & 0.845 & 4.41 & 0.57 & 4.14 & 0.20 & 1.147 \\
2459011.26 & 0.048 & -34.34 & 0.58 & -34.28 & 0.34 & 1.686 \\
2459011.27 & 0.052 & -34.75 & 0.91 & -35.92 & 0.27 & 2.194 \\
2459011.29 & 0.056 & -37.30 & 0.94 & -37.64 & 0.61 & 2.304 \\
2459011.30 & 0.060 & -38.18 & 0.94 & -38.57 & 0.66 & 2.641 \\
2459015.20 & 0.004 & -27.25 & 0.29 & -26.35 & 0.11 & 0.958 \\
2459015.22 & 0.008 & -27.55 & 0.31 & -27.33 & 0.14 & 1.035 \\
2459015.23 & 0.012 & -28.89 & 0.31 & -28.51 & 0.14 & 1.013 \\
2459015.25 & 0.018 & -29.54 & 0.41 & -30.02 & 0.17 & 0.988 \\
2459015.27 & 0.022 & -31.01 & 0.32 & -31.16 & 0.16 & 1.005 \\
2459015.29 & 0.026 & -31.88 & 0.61 & -31.99 & 0.14 & 1.016 \\
2459016.24 & 0.257 & -59.67 & 0.43 & -58.18 & 0.19 & 2.378 \\
2459016.26 & 0.261 & -58.91 & 0.75 & -57.85 & 0.20 & 2.364 \\
2459016.27 & 0.265 & -57.68 & 0.98 & -58.20 & 0.26 & 2.288 \\
2459016.29 & 0.269 & -59.35 & 0.83 & -57.62 & 0.24 & 2.334 \\

\end{longtable}

% -----------------------------
\clearpage
\section{Markov Chain Monte Carlo sampling for spot modeling}\label{sec:MCMC}
\begin{figure*}[ht!]
\centering
        \includegraphics[width=0.85\linewidth]{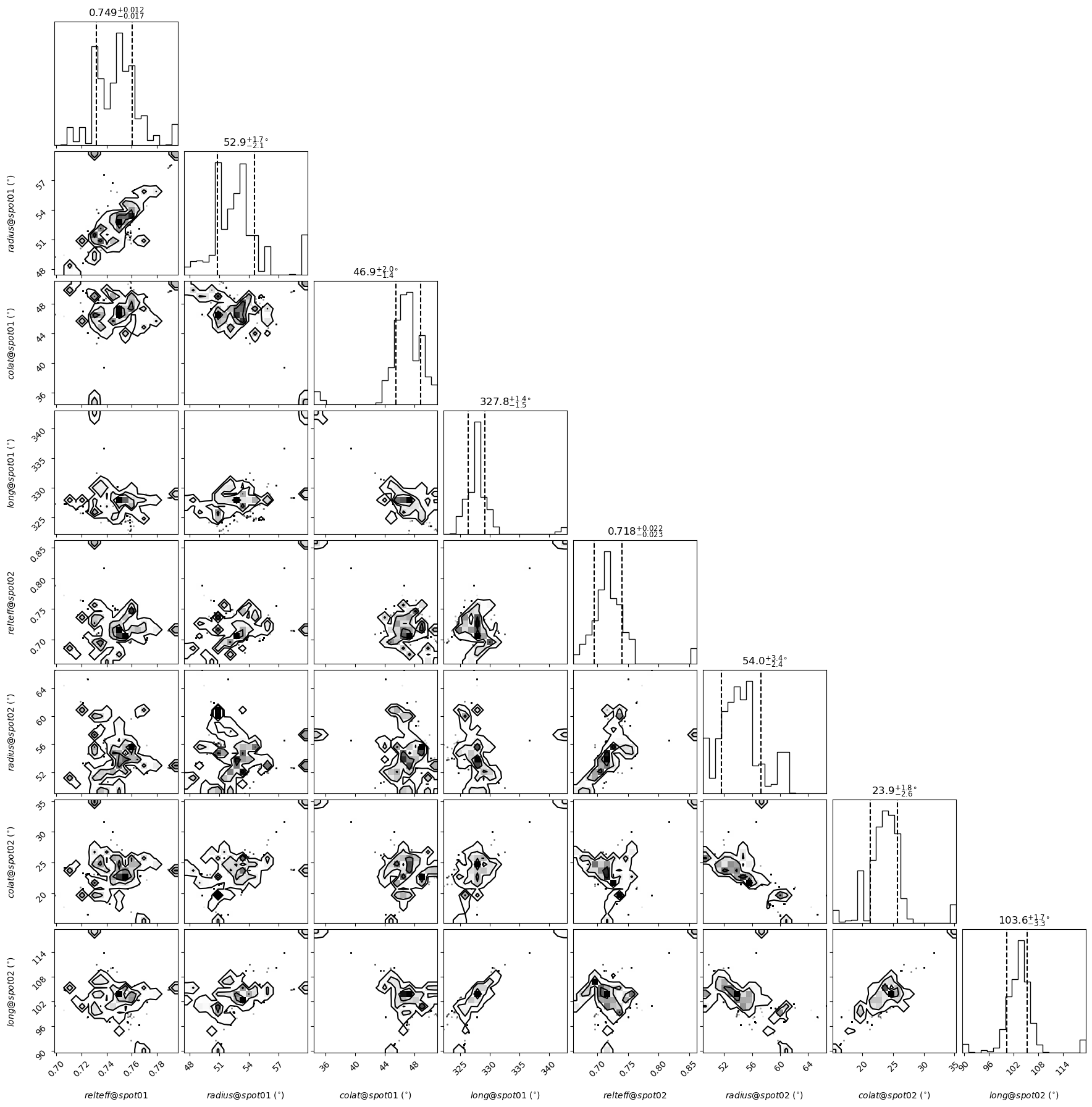}
    \caption{Corner map of the MCMC fitting result. The dashed lines show the 1$\sigma$ edges of the MCMC results.}
    \label{fig:MCMC}
\end{figure*}

% -----------------------------
\clearpage
\section{Flare Energies\label{sec:tessflare}}
\begin{figure}[ht!]
\plotone{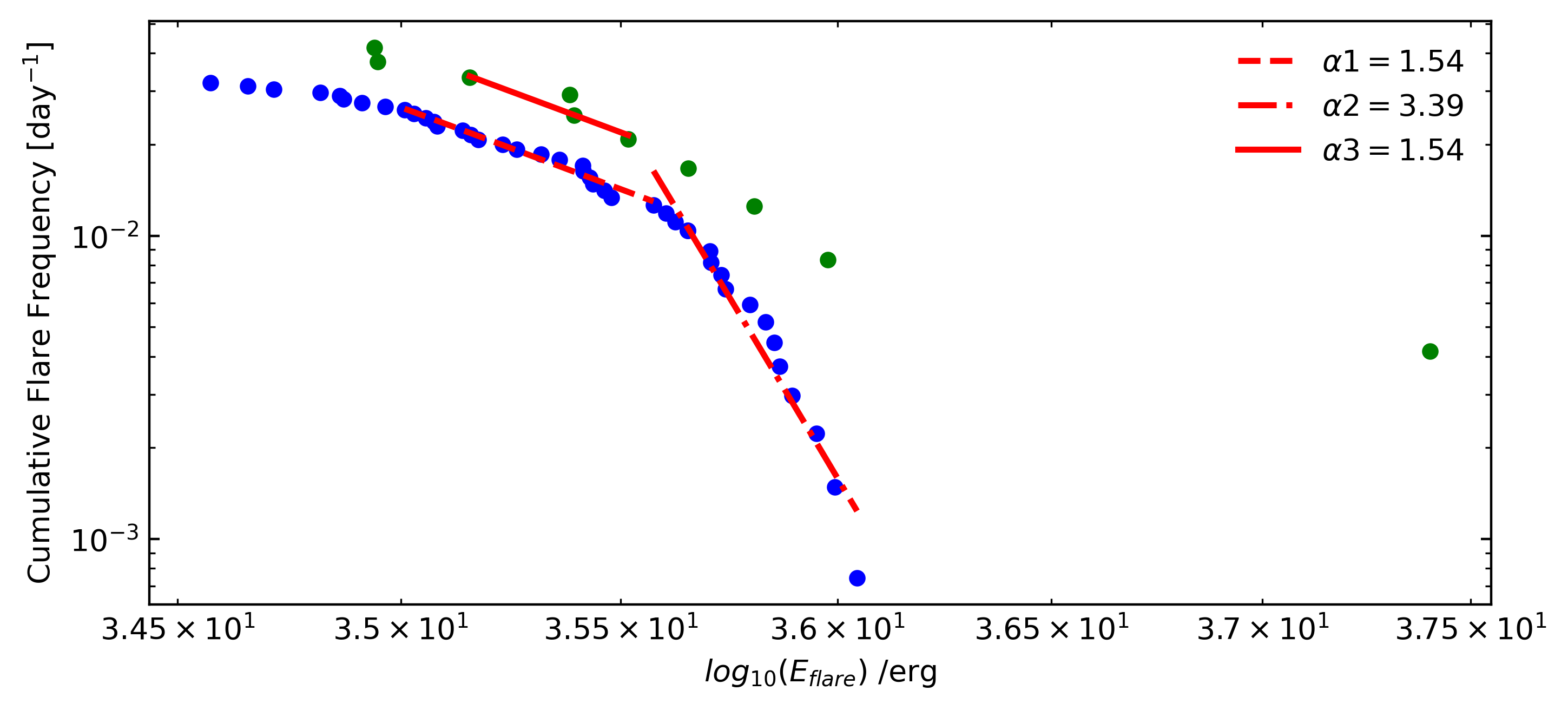}
\caption{The corresponding cumulative flare frequency distribution (FFD) of \textit{Kepler} (blue) and \textit{TESS} (green) flares. The $\alpha$ parameters of Eq.\ref{eq:ffd} fitting are shown with red lines.
\label{fig:flare_ffd}}
\end{figure}

We investigated the flare frequency distribution (FFD) using the cumulative FFD formalism \citep{1976ApJS...30...85L,2024ApJS..271...57X}, expressed as:

\begin{equation}
\label{eq:ffd}
\mathrm{log}(\nu) = \beta + (1 - \alpha)\mathrm{log}(E),
\end{equation}

where $\nu$ is the number of flares per year with an energy above a specific threshold, and $E$ represents the energy of the flares. The phase-folded distribution of flare energies is presented in Fig. \ref{fig:flare}, and the corresponding cumulative FFD is shown in Fig. \ref{fig:flare_ffd}. The fitted $\alpha$ for V2279 Cyg is $1.54\pm0.2$ for both \textit{Kepler} and \textit{TESS}, which is similar to the $\alpha=1.52$ for PMS binary DQ Tau \citep{2018ApJ...862...44K}.

\begin{table}[ht!]
\centering
\caption{The energies of flares in \textit{TESS}}
\begin{tabular}{ccccll}
\cline{1-4}
\textbf{$\mathrm{Time\_Start}$} & \textbf{$\mathrm{Time\_Stop}$} & \textbf{$E_{flare}$}    & \textbf{$\mathrm{Sector}$} &  &  \\ (BKJD) & (BKJD) & (erg) & & & \\ \cline{1-4}
3854.572184        & 3855.280526       & 2.53E+37 & 14     &  &  \\
3860.738926        & 3860.947262       & 4.51E+35 & 14     &  &  \\
4573.874051        & 4573.943496       & 8.87E+34 & 40     &  &  \\
4574.401833        & 4574.575446       & 1.43E+35 & 40     &  &  \\
4594.929678        & 4595.138011       & 6.41E+35 & 41     &  &  \\
4604.089338        & 4604.242114       & 2.47E+35 & 41     &  &  \\
4954.519795        & 4954.867017       & 9.49E+35 & 54     &  &  \\
5488.110207        & 5488.161133       & 8.72E+34 & 74     &  &  \\
5501.994595        & 5502.119598       & 2.42E+35 & 74     &  &  \\
5528.301021        & 5528.395933       & 3.29E+35 & 75     &  &  \\ \cline{1-4}
\end{tabular}
\end{table}

\begin{figure}[ht!]
\plotone{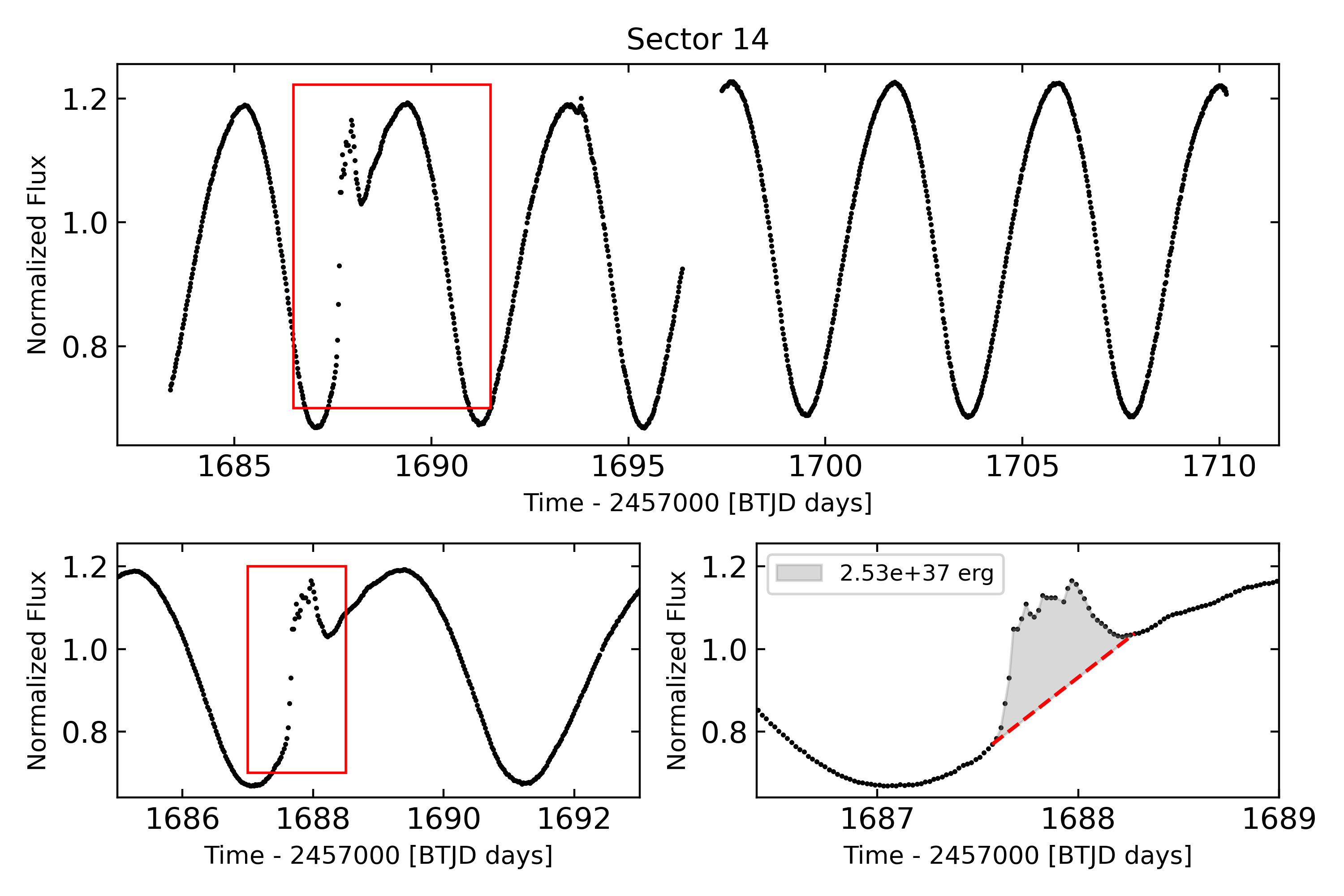}
\caption{The superflare in \textit{TESS} Sector 14.
\label{fig:superflare}}
\end{figure}

\clearpage
\section{SED}
\label{sec:sed}
\begin{figure}[ht!]
\centering
        \includegraphics[width=0.45\linewidth]{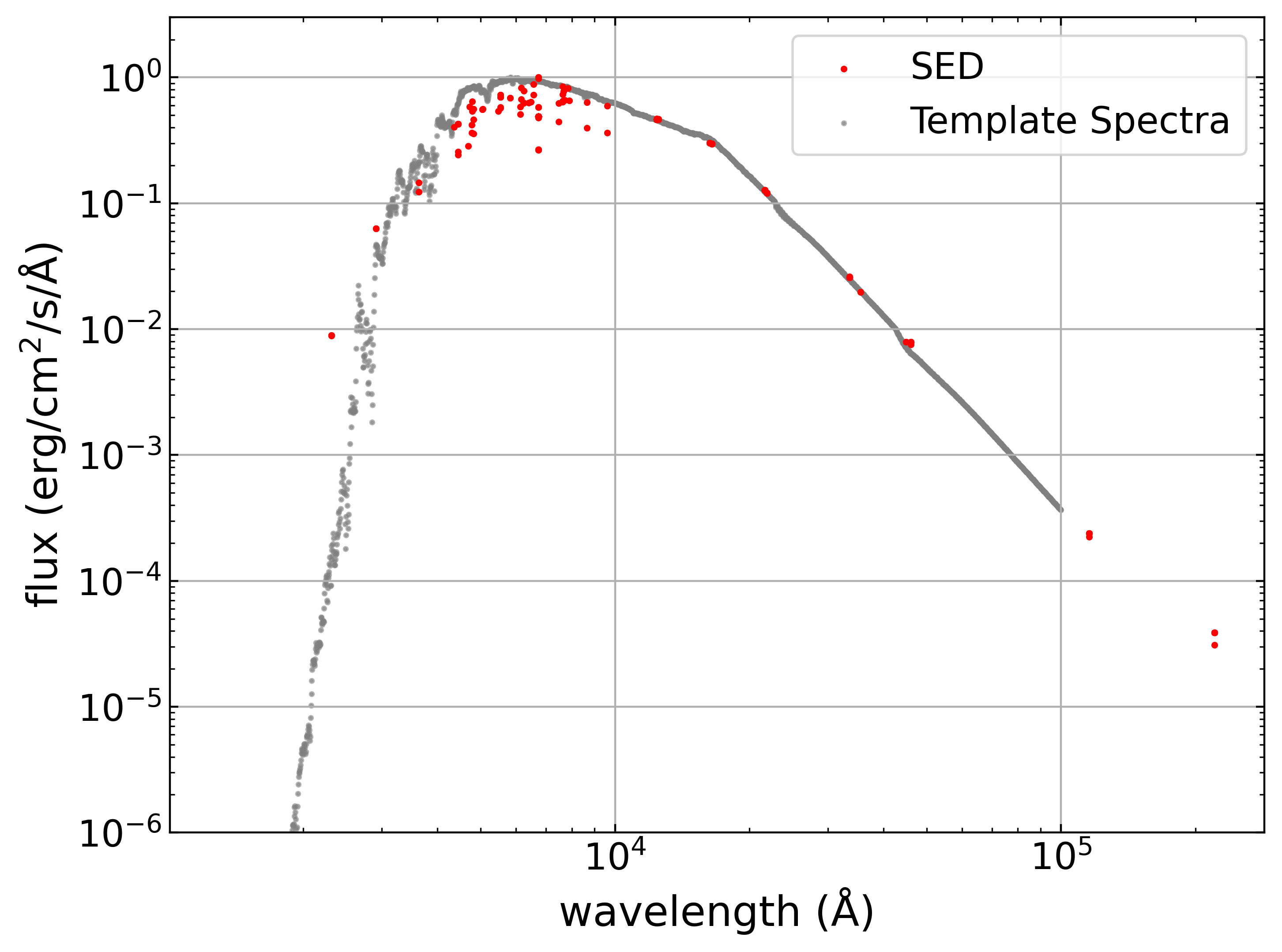}
    \caption{SED of V2279 Cyg. The template spectra is from \cite{2014MNRAS.440.1027C}, with $T_{\mathrm{eff}} = 4500$ K, log $g$ = 3, [Fe/H] = -0.5, [$\alpha$/Fe] = 0. The SED data were from VizieR \citep{10.26093/cds/vizier}.}
    \label{fig:SED}
\end{figure}

% -----------------------------
\clearpage
\section{LSD and H-alpha profiles}
\label{sec:ha}

\begin{figure}[ht!]
    \centering        \includegraphics[width=0.5\linewidth]{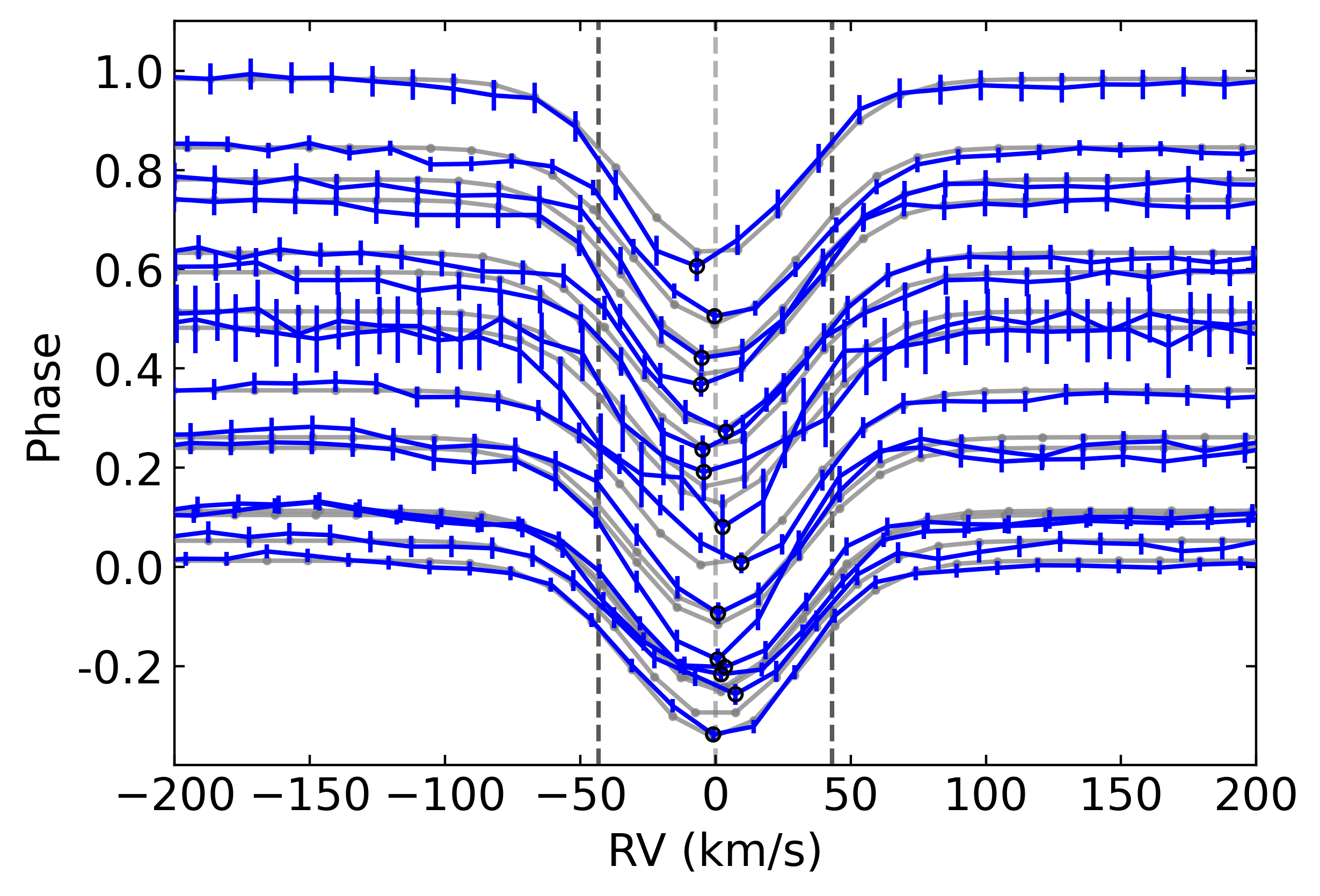}
    \caption{LSD profiles (blue lines with error bars) arranged by phase, spanning from 2018 to 2020. The gray lines represent the mean photospheric structures modeled using Gaussian profiles. Black circles indicate the minimum value of each profile. The vertical black dashed lines mark the $\pm v \mathrm{sin}i$.}
    \label{fig:lsd_profile}
\end{figure}

\begin{figure*}[h]
    \centering
    \includegraphics[width=0.45\linewidth]{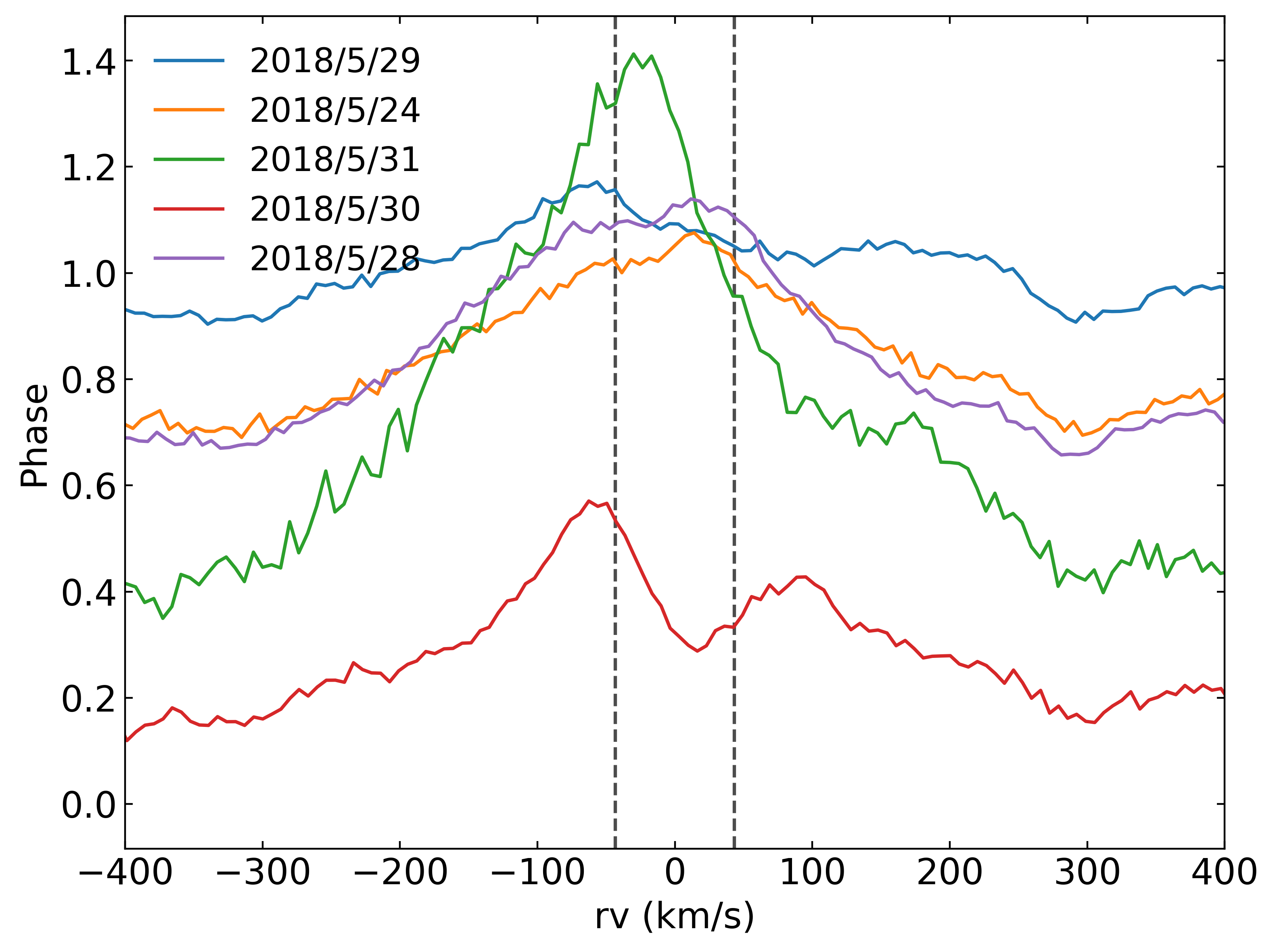}%2018_line.png
    \includegraphics[width=0.45\linewidth]{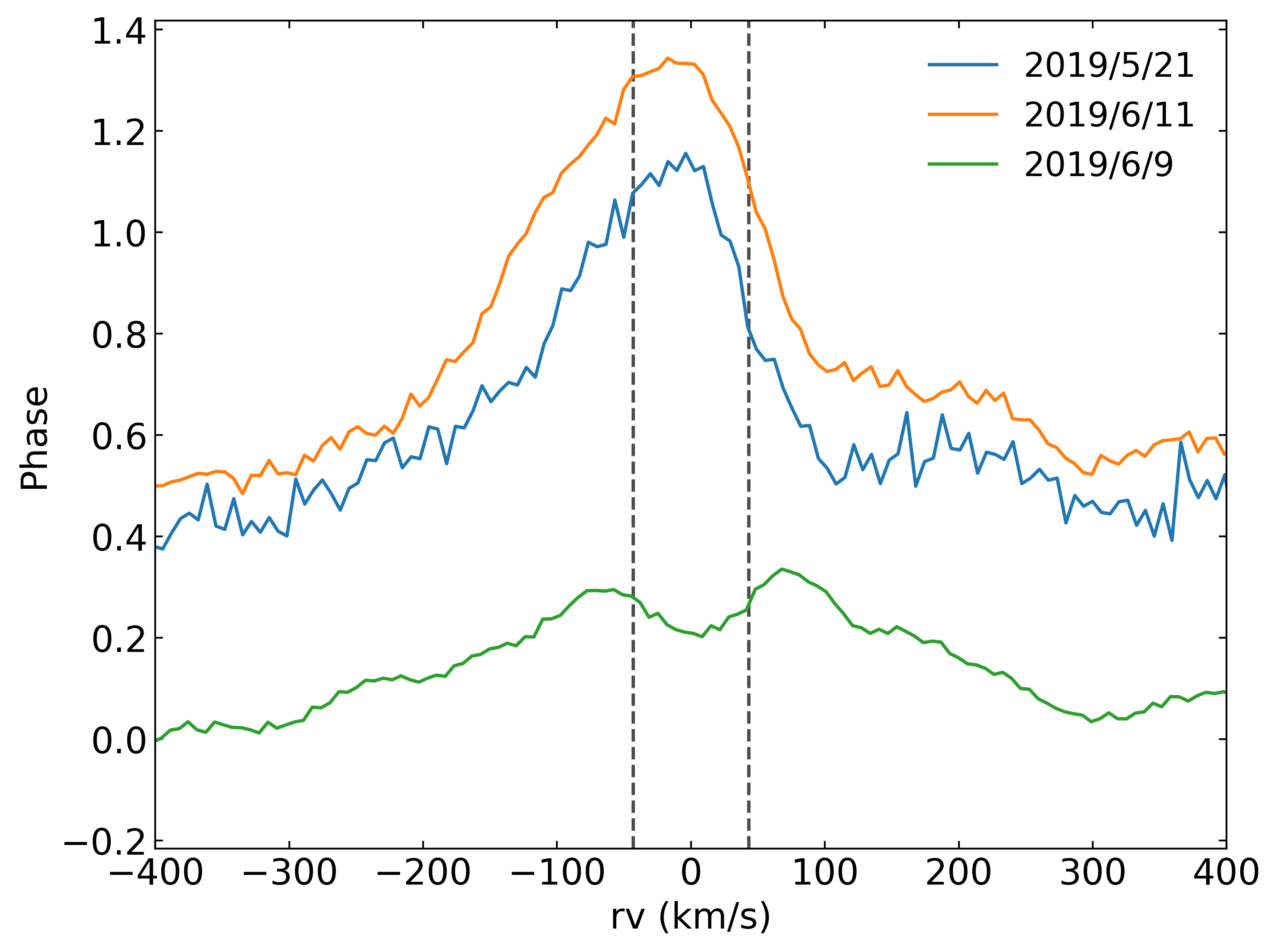}
    \includegraphics[width=0.45\linewidth]{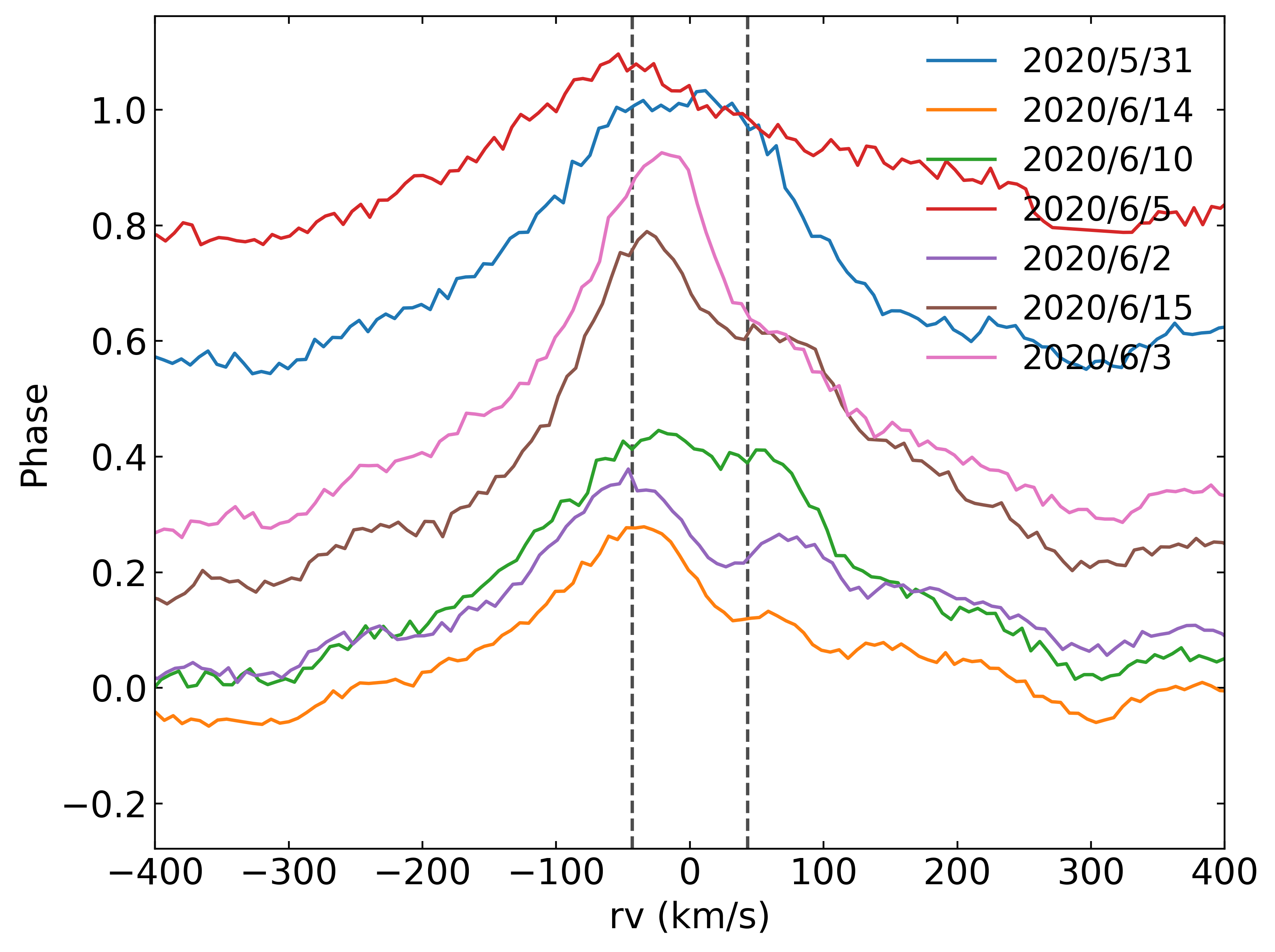}
    \caption{H$\alpha$ profiles averaged over individual nights during 2018-2020 (from left to right), after correcting the RV caused by the binary system. $\mathrm{BJD_0} = 2458268.3132$ was chosen as the initial time for phase-folding. Two vertical dashed lines show the $\pm v \mathrm{sin}i$ \citep{2022A&A...664A..78F}.}
    \label{fig:ha_line}
\end{figure*}

To enhance the S/N of LAMOST spectra, we applied the LSD technique, and the resulting profiles are presented in Fig. \ref{fig:lsd_profile}. Variations induced by stellar spots are detectable. Additionally, the original spectra reveal prominent H$\alpha$ emissions, which exhibit temporal variability (Fig. \ref{fig:ha_line}).

\bibliography{references}{}
\bibliographystyle{aasjournal}

\end{document}